\documentclass[a4paper,11pt]{article}
\pdfoutput=1
\usepackage{jcappub}
\usepackage[T1]{fontenc}
\usepackage{graphicx}
\usepackage{amsmath,amssymb}
\usepackage{mathcomp}
\usepackage{url}
\usepackage{epstopdf}
\allowdisplaybreaks
\numberwithin{equation}{section}
\usepackage{enumitem}
\usepackage{bm}
\usepackage[utf8]{inputenc}
\usepackage{color}
\usepackage{bm}
\usepackage{multirow}
\usepackage{hyperref}
\usepackage{multirow}
\usepackage[dvipsnames]{xcolor}

\usepackage{subfigure}
\usepackage{array}
\usepackage{todonotes}
\usepackage[english]{babel}
\usepackage{tensor}
\usepackage{comment}
\usepackage[normalem]{ulem}

\hypersetup{
    colorlinks=true,
    linkcolor=blue,
   citecolor=blue,
   urlcolor=blue
}
\graphicspath{{Figures/}}

\definecolor{abpurple}{rgb}{0.75, 0.0, 1.0}

\newcommand{\abc}[1]{\textcolor{abpurple}{[AB: #1]}}

\newcommand{\lcdm}{$\Lambda$CDM}
\def\H{\mathcal{H}}
\def\vpar{v_{\parallel}}
\def\bchi{\bar{\chi}}
\def\hchi{\bar{\chi} \, \mathcal{H}}
\def\btheta{\bar{\theta}}
\def\bphi{\bar{\phi}}
\def\bn{\bar{n}}
\def\bD{\bar{D}}
\def\bz{\bar{z}}

\newcommand{\sprod}[2]{\bm{#1}\cdot\bm{#2}}
\usepackage{orcidlink}
\title{Number count of Gravitational Waves and Supernovae in Luminosity Distance space for \lcdm{} and Scalar-Tensor theories}

\author[a,b]{Anna Balaudo\orcidlink{0000-0003-4109-8094},}
\emailAdd{balaudo@strw.leidenuniv.nl}

\author[b]{Mattia Pantiri\orcidlink{0009-0007-2259-2368},}
\author[b]{Alessandra Silvestri\orcidlink{0000-0001-6904-5061}}
\affiliation[a]{Leiden Observatory, Leiden University, PO Box 9513, Leiden 2300 RA, The Netherlands}
\affiliation[b]{Institute Lorentz, Leiden University, PO Box 9506, Leiden 2300 RA, The Netherlands}

\abstract{
The clustering of gravitational waves in luminosity distance space is emerging as a promising probe of the growth of structure. Just like for galaxies, its observation  is subject to a number of relativistic corrections that affect the measured signal and need to be accounted for when fitting theoretical models to the data. We derive the full expression for the number count of gravitational waves in luminosity distance space, including all relativistic corrections, in $\Lambda$CDM and in scalar-tensor theories with luminal propagation of tensors. We investigate the importance of each relativistic effect and the detectability of the total signal by current and planned GW detectors. We consider also supernovae in luminosity distance space, highlighting the differences with gravitational waves in the case of scalar-tensor theories. We carry out a thorough comparison among the number count of gravitational waves and  supernovae in luminosity distance space, and that of galaxies in redshift space. We show how the relativistic corrections contain useful complementary information on the growth of perturbations and on the underlying theory of gravity, highlighting the synergy with other cosmological probes.}

\begin{document}

\maketitle
\flushbottom

\section{Introduction}

The direct detection of gravitational waves (GW)~\cite{LIGOScientific:2016aoc} has opened a new window onto the Universe, that can be highly complementary to EM observations. The use of GW as probes of cosmology was first advanced in the '80s by Schutz~\cite{Schutz:1986gp}, who proposed the standard sirens method to measure the present value of the Hubble constant. This became a reality in 2017, when LIGO-Virgo-Kagra (LVK) detected the first merger of two binary neutron stars, GW170817~\cite{LIGOScientific:2017vwq}, while a network of EM detectors was able to measure the coincident EM emission from the merger and remnant~\cite{LIGOScientific:2017ync}, placing the value of $H_0$ at $70.0^{+12.0}_{-8.0}$~\cite{LIGOScientific:2017adf}. However, events such as GW170817 are a lucky rarity, while most of the GW detected in LVK are not accompanied by a EM counterpart, nor they are expected to be in the future.
Moreover, standard sirens also present a big limitation, in that they are only sensitive to the subset of cosmological parameters that determines the background expansion. At the same time, they are totally unaffected by other important parameters that drive the growth of structure. To efficiently test the full extent of the cosmological model then, other methodologies must be investigated.

One possibility that took hold in recent years is to use GWs as tracers of the large scale structure (LSS), analogously to what is normally done with galaxies. This builds on the reasonable assumption that, with the possible exception of exotic sources such as primordial black holes, the binaries that generate GWs are located within or in the proximity of galaxies, and share similar clustering properties. On top of that, the GWs that reach the detector have travelled through a Universe rich of small, local perturbations, which have imprinted on the wave by slightly altering its shape~\cite{Laguna:2009re, Hirata:2010phr, Bertacca:2017vod, Garoffolo:2019mna}. There are two complementary roads to approaching LSS studies with GW: the weak lensing convergence field can be probed by measuring perturbations induced on the amplitude of the incoming wave, as explored e.g. in~\cite{Camera:2013xfa, Mukherjee:2019wcg, Mukherjee:2019wfw, Congedo:2018wfn, Mpetha:2022xqo, Balaudo:2022znx}; or, the number count, or clustering, of compact objects - defined as the local abundance in the number of detected GW sources in the sky - which can by itself constitute the observable. This second possibility has been explored before in~\cite{Namikawa:2015prh, Namikawa:2020twf, Libanore:2020fim, Libanore:2021jqv, Scelfo:2021fqe, Scelfo:2022lsx, Fonseca:2023uay}, and it is also the focus of this work.

It is well known that, when performing clustering studies with galaxies, the signal is affected by relativistic effects due to the fact that the number density of galaxies is measured in redshift bins, whose actual central value and volume are distorted by perturbations locally and along the line of sight. These effects have first been derived in the linear regime in~\cite{Tadros:1999ky, Samushia:2012mnr, Yoo:2009cfg, Yoo:2010ni, Challinor2011, Bonvin2011, Jeong:2012rvd}.
Similar caution must be applied in the case of GWs, with the crucial difference that the number density is now measured in luminosity distance space (or D-space). In fact, except for few cases in which the redshift of the source can be determined, GWs are dark standard sirens. The inclusion of the relativistic effects for the clustering of GWs in luminosity distance space has been explored in~\cite{Namikawa:2020twf, Libanore:2020fim,Libanore:2021jqv, Yang:2022uye}, where lensing  and luminosity distance space distortions where taken into account. More recently, ~\cite{Fonseca:2023uay} provided a full theoretical expression of the number count of GWs in luminosity distance space, including all corrections  for a \lcdm{} Universe. This work retraced the calculation performed in redshift space (or z-space) by~\cite{Bonvin2011}, connecting the GWs number count to the luminosity distance perturbations of photons and GWs derived  in~\cite{Sasaki:1987, Pyne:2003bn, Hui:2005nm, Bacon:2014uja, Bertacca:2017vod}. In this paper, we build on~\cite{Garoffolo:2019mna, Bertacca:2017vod} and~\cite{Fonseca:2023uay} by further extending the calculation to the broader class of scalar-tensor theories of gravity with luminal speed of sound for the propagating tensors.

The paper is organized as follows. In  Sec.~\ref{sec:reference_frames} we re-derive the result of~\cite{Fonseca:2023uay} in a more general form, without ever making any assumption about the background luminosity distance-redshift relation, so that it can be readily generalized to theories beyond GR. In Sec.~\ref{sec:number_count_st} we specialize scalar-tensor (ST) theories, writing the full expression of the luminosity distance number count for the sub-set of Horndeski theories with $c_T=1$.
We caution that, in the GR limit, our result reproduces that of~\cite{Fonseca:2023uay}, up to a minus sign in the coefficients of six relativistic corrections. We comment more about this in the text. In Sec.~\ref{sec:correlations}, we extend our result to supernovae (SN). The latter also live naturally in luminosity distance space, however, in the case of ST, they do not receive any explicit corrections from the modifications of the theory (see e.g.~\cite{Garoffolo:2019mna}). Thus, the number count of SN in luminosity distance space will have a different form than that of GWs. We provide both expressions and comment on the implications for testing gravity theories in Sec.~\ref{sec:correlations}. In Sec.~\ref{sec:correlations}, we also build the 2-point angular correlation functions of the luminosity distance number count for GWs and SN, and  compare them with the case of galaxies (in redshift space), exploring the potentialities of a joint cosmological analysis. We investigate the weight that the individual relativistic components have on the total signal, indicating under which circumstances  they can be safely neglected.
Finally, in Sec.~\ref{sec:observations} we assess the potentiality for observation of the number count correlation of compact objects, by plotting its cumulative signal-to-noise ratio for a set of different benchmark scenarios, considering 3rd and 4th generation GW detectors.
We draw our conclusions in Sec.~\ref{sec:conclusions}.

\section{Number Count in Luminosity Distance Space}
\label{sec:reference_frames}

Throughout our derivation for the number count in luminosity distance space, we will distinguish between two different reference frames: the smooth, \textit{background}, frame and the observed, or \textit{perturbed}, frame. This distinction is necessary because the local inhomogeneities in the Universe perturb the propagation of photons and gravitational waves~\cite{Sasaki:1987, Bertacca:2017vod}. The direct consequence is that all astrophysical parameters inferred by measuring the perturbed waves are no longer the true, physical ones, but will differ by a small amount.

More rigorously, the perturbed frame contains all measurable quantities. In this frame, the past comoving lightcone of GWs (or photons) is not flat, as the wave propagation is perturbed by relativistic effects induced by the large scale structures. We use un-barred comoving coordinates $(\eta, \bm{x})$, and we describe the past lightcone in terms of the corresponding comoving distance $\chi$. In the observed frame, every celestial object has a redshift $z(\chi)$ and a corresponding luminosity distance $D(\chi)$. Both quantities are the ones that we can infer directly from data, modulus the measurement uncertainty.
 
At the same time, each object also exists in the background frame. Here, it has a comoving position identified by barred coordinates $(\bar{\eta}, \bar{\bm{x}})$ and its past lightcone, parametrized in terms of the true comoving distance $\bchi$, is flat, so that the background comoving geodesic has coordinates
\begin{equation}
    \bar{x}^{\mu} = (\eta_0-\bchi, \hat{n}  \bchi)\,.
\end{equation}
Here, $\hat{n}$ is the direction of arrival of the incoming wave, i.e. the unit vector is oriented from the observer to the source, $\hat{n}^i\equiv \bar{x}^i/\bchi$. This means that the total derivative along the wave geodesic can be computed as\footnote{Note that we use the opposite convention for $\hat{n}$ than what is done in~\cite{Bonvin2011}, while we align to~\cite{Fonseca:2023uay}.} 
\begin{equation}
 \frac{d}{d\bchi} = -\frac{\partial}{\partial \bar{\eta}} + \hat{n}^i \frac{\partial}{\partial \bar{x}^i}\,.
\end{equation}

In this frame, the object has a background redshift $\bar{z}(\bchi)$, with corresponding luminosity distance $\bar{D}(\bchi)$. These can be thought of as the true redshift and luminosity distance of the object or, analogously, the redshift and distance that would be inferred through measurements of photons and gravitational waves if the Universe was perfectly homogeneous on all scales. We now define the quantity $\Delta D$ as the difference between the luminosity distance in the perturbed frame and the one in the background frame, i.e.
\begin{equation} \label{eq:delta_D_chi}
    \Delta D (\chi, \hat{n}) \equiv D(\chi, \hat{n}) - \bar{D}(\bchi) = \delta D (\bchi, \hat{n}) + \frac{d D}{d\chi}\bigg|_{\chi=\bchi} \delta \chi (\hat{n})\,,
\end{equation}
where we have further defined $\delta D (z, \hat{n}) \equiv D(z(\chi), \hat{n}) - \bar{D}(z(\bchi))$. In other words, we can split the luminosity distance perturbations in two contributions: $\delta D$ is the effect resulting from the perturbation of the past lightcone, which introduces a difference between the observed and physical luminosity distance even when these are evaluated at the same value of the affine parameter; the term proportional to $\delta \chi$ accounts instead for the variation of the affine parameter itself in going from one frame to the other. Note that for few terms in Eq.~\eqref{eq:delta_D_chi} we introduced  a dependency on the specific line of sight $\hat{n}$. This is because the perturbed luminosity distance observed for each object depends on the specific configuration of the inhomogeneities inducing the wave perturbations, both at the source location (as is the case for perturbations induced by the local potentials and peculiar velocities) and along the wave path (as is the case for integrated effects, such as lensing). Eq.~\eqref{eq:delta_D_chi} can also be re-written in terms of the observed redshift $z(\chi)$ and the true redshift $\bz(\bchi)$ as
\begin{equation} \label{eq:delta_D}
    \Delta D [\bz(\bchi), \hat{n}] \equiv D[z(\chi), \hat{n}] - \bar{D}[\bz(\bchi)] \, .
\end{equation}

The mapping of the observed universe onto the smooth background Friedmann universe is necessary, if one wishes to exploit all standard cosmological relations connecting the background quantities such as $\bD$, $\bz$, $\bchi$, $\bar{\eta}$ etc. However, in general it introduces a gauge dependence for the perturbations, as they will now depend on the specific choice made for the background, which is known as the cosmological gauge problem~\cite{Bardeen:1882, Durrer:1993tti}. This gauge dependence will cancel out in the final expression of the observed number count defined below (see Eq.~\eqref{eq:observable_clustering}), as $\hat{\Delta}$ is built fully from observable quantities~\cite{Bonvin2011}. 

Having made this distinction between frames, we can now give a formal definition for the number count, following the standard definition used for galaxies, translated in luminosity distance space~\cite{Namikawa:2015prh, Namikawa:2020twf, Libanore:2021jqv, Fonseca:2023uay}
\begin{equation} \label{eq:observable_clustering}
    \hat{\Delta}(D, \hat{n}) \equiv \frac{N(D, \hat{n}) - \bar{N} (D)}{\bar{N}(D)}\,,
\end{equation}
where $N(D, \hat{n})$ is the number of objects detected within a small volume at observed luminosity distance $D$ and in direction $\hat{n}$, and $\bar{N}(D)$ is its angular average over the whole sky. As it is built from observational data, $N$ lives in the perturbed frame. On the contrary, $\bar{N}(D)$ is an ambiguous object, as its explicit dependence on the specific perturbations of the wave lightcone (i.e. on $\hat{n}$) has been removed by the average, but it is still evaluated at the perturbed luminosity distance. Thus, to fully work out the theoretical expression for $\hat{\Delta}$, we need to further expand $\bar{N}$ and refer it to background quantities.

Before doing so, we follow~\cite{Bonvin2011} and use that $N \equiv \rho V$, where $\rho$ is the number density of events, and $V$ is the small sky-volume considered. In terms of averaged quantities, we also have $ \bar{N}(D) = \bar{\rho}(D) \bar{V}(D)$, and we can relate averaged and un-averaged as
\begin{equation}\begin{split}\label{eq:density_expansion}
    \rho(D, \hat{n}) &= \bar{\rho}(D) + \delta\rho(D, \hat{n})\,, \\
    V(D, \hat{n}) &= \bar{V}(D) + \delta V(D, \hat{n})\,,
\end{split}
\end{equation}
analogously to what we did in Eq.~\eqref{eq:delta_D_chi}. Here $\delta \rho$ and $\delta V$ are assumed to be small. Inserting Eqs.~\eqref{eq:density_expansion} in~\eqref{eq:observable_clustering} and keeping only linear order terms in the perturbations, we obtain
\begin{equation} \label{eq:splitted_clustering}
    \hat{\Delta} (D, \hat{n}) = \frac{\rho(D) - \bar{\rho}(D)}{\bar{\rho}(D)} + \frac{V(D) - \bar{V}(D)}{\bar{V}(D)}\,,
\end{equation}
i.e. we can split the number count into the sum of two separate perturbations, one on the number density and one on the volume. We can now further expand the barred quantities at the RHS of Eq.~\eqref{eq:splitted_clustering}, and refer them to the background distance, obtaining, at linear order
\begin{equation} \label{eq:counts_expanded}
    \hat{\Delta} (D, \hat{n}) = \frac{\rho(D) - \bar{\rho}(\bar{D})}{\bar{\rho}(\bar{D})} - \frac{1}{\bar{\rho}(\bar{D})} \frac{\partial \bar{\rho}}{\partial D}\bigg|_{\bar{D}} \Delta D + \frac{V(D) - \bar{V}(\bar{D})}{\bar{V}(\bar{D})} - \frac{1}{\bar{V}(\bar{D})} \frac{\partial \bar{V}}{\partial D}\bigg|_{\bar{D}} \Delta D\,.
\end{equation}

The first term in Eq.~\eqref{eq:counts_expanded} is the standard definition of the density contrast, or clustering, of gravitational waves $\delta^N_{\rm gw}$ in Newtonian gauge. The same quantity in synchronous gauge can be linked to the density contrast of the underlying dark matter distribution through a bias parameter $b_{\rm gw}$ so that $\delta^S_{\rm gw} = b_{\rm gw} \delta^S_m$. The calculation of the three remaining terms can be pushed forward explicitly, keeping into account the dependence of the background density and volume on the background distance. We do so in App.~\ref{app:density_perturbations} for the density term and in App.~\ref{app:volume_perturbations} for the volume terms. In doing so, we follow closely the calculation made in \lcdm{} by~\cite{Bonvin2011}, that was recently adapted to the luminosity distance case by~\cite{Fonseca:2023uay}. However, we maintain an extra degree of generality in our calculations, so that we can then proceed to extend the result to scalar-tensor theories of gravity. In particular, we first compute the RHS of Eq.~\eqref{eq:counts_expanded} without ever substituting i) the explicit expression of the luminosity distance perturbations $\Delta D$ and ii) the explicit dependence of the background luminosity distance $\bar{D}$ and its first and second derivatives on the background redshift $\bar{z}$.

We also work in Newtonian gauge, for which the metric perturbations in the perturbed frame are
\begin{equation}\label{eq:perturbed_metric}
    ds^2 = a^2(\eta) [-(1+2\Psi) d\eta^2 +(1-2\Phi) \delta_{ij} dx^i dx^j]\,.
\end{equation}

Defining
\begin{equation}\label{eq:gamma_def}
    \gamma \equiv (a \bar{D}) \bigg(\frac{d\bar{D}}{d\bar{z}}\bigg)^{-1}\,,
\end{equation}
our result for the luminosity distance number count in its general form is\footnote{We caution that, in generalizing this result, we spotted a typo in Eq. (2.63) in~\cite{Fonseca:2023uay}: where in the last squared parenthesis the last term for us is a $+1$, they instead have $-1$. This simple difference has quite an impact on the results, as we will show in the next sections.}
\begin{equation}\begin{split}\label{eq:counts_general}
    \hat{\Delta} (D, \hat{n}) = &\delta^N_{\rm gw} + \frac{2}{\bchi{}} \int_0^{\bchi{}} d\chi' (\Phi+\Psi) - \int_0^{\bchi{}} d\chi' \frac{\bchi{}-\chi'}{\bchi{}\chi'} \nabla_{\Omega}^2 (\Phi + \Psi)  + \bm{v}\cdot\hat{n} - 2\Phi + \Psi + \\
    & - \frac{\gamma}{\H{}}\frac{d}{d\bchi} \bigg(\frac{\Delta D}{\bar{D}}\bigg)  - \bigg[\gamma \bigg(\frac{2}{\bchi{} \H{}} + \frac{\H{}'}{\H{}^2} - \frac{\gamma}{a^2 \bar{D}} \frac{d^2\bar{D}}{d\bar{z}^2} -1 - \frac{\partial \ln \bar{n}}{\partial \ln a} \bigg)+1 \bigg] \frac{\Delta D}{\bar{D}}\,.
\end{split}
\end{equation}

In the first line of Eq.~\eqref{eq:counts_general} we can already recognise a number of perturbative terms that have similar origin to those affecting the number count of galaxies. In order, these are the density contrast term, the time delay term, a lensing term, the Doppler perturbation and the perturbations due to the local potentials at the source position. More perturbative terms are hidden in the second line of the equation, and will become explicit once we write the full expression for the luminosity distance perturbations in the desired MG theories, and compute their derivatives along geodesics.

\section{GW number count in scalar-tensor theories of gravity}
\label{sec:number_count_st}

We consider the class of scalar-tensor theories  described by the following action:
\begin{equation}\label{eq:action}
    S = \int d^4x \sqrt{-g} \bigg[F(\varphi) R + G(\varphi, X) \Box \varphi + K(\varphi, X) \bigg]\,,
\end{equation}
where $\varphi$ is the additional scalar degree of freedom,  $X$ its kinetic term, $X= - \partial_{\mu} \varphi \partial^{\mu} \varphi /2$, $F(\varphi)$ represents a non-minimal coupling and $G, K$ are general functions of $\varphi$ and $X$ which can lead to kinetic braiding and non-trivial kineticity of the scalar field. Action~(\ref{eq:action}) corresponds to the subset of the broader Horndeski class of theories \cite{Horndeski:1974wa} that leads to tensors propagating at the speed of light. This requirement is inspired by  the LIGO observations of the binary neutron star system GW170817~\cite{LIGOScientific:2017zic}, which placed significant limitations to the class of surviving MG theories, as investigated in~\cite{Ezquiaga:2017ekz,Creminelli:2017sry,Baker:2017hug} (see also~\cite{deRham:2018red} for a discussion of caveats). More concretely, the $c_T^2=1$ restriction is dictated by the fact that relativistic corrections to the luminosity distance of GWs have been calculated only for Horndeski theories with $c_T^2=1$~\cite{Garoffolo:2019mna}. In~\cite{Garoffolo:2020vtd}, some of the authors generalized the calculation to DHOST~\cite{Langlois:2015cwa} theories with $c_T^2=1$. It would be straightforward to apply our calculation to this broader class, however the difference is negligible in terms of impacts on the number count and, for simplicity, in this work we focus on the Horndeski case.

We use the following convention for splitting the scalar field into a  background and perturbation component:
\begin{equation}\label{eq:scalar_field_perturbations}
    \varphi(\chi, \hat{n}) = \bar{\varphi} (\bar{\chi}) + \Delta \varphi(\bar{\chi}, \hat{n}) \quad, \qquad \text{with} \qquad \Delta \varphi (\bar{\chi}, \hat{n}) = \delta \varphi(\bar{\chi}, \hat{n}) + \delta \chi \frac{d\bar{\varphi}}{d\bchi} \bigg|_{\bchi}\,,
\end{equation}
where again, $\delta$ represents variations of the field evaluated at the same affine parameter, while $\Delta$ describes the full perturbation. We stress here that the perturbations of the scalar field, similarly to $\Psi$ and $\Phi$ for the perturbed metric in~\eqref{eq:perturbed_metric}, are considered to be long wavelength perturbations, clearly distinguished from the high frequency gravitational waves. The coefficient functions in the EFT action are perturbed accordingly. In particular, the function $F(\varphi)$ that introduces the non-minimal coupling can be expanded as
\begin{equation}
    F(\varphi) = \bar{F}(\bar{\varphi}) \bigg[ 1 + \frac{\delta F(\bar{\varphi})}{\bar{F}(\bar{\varphi})} + \bigg( \delta \varphi(\bar{\chi}, \hat{n}) + \delta \chi \frac{d\bar{\varphi}}{d\bchi} \bigg|_{\bchi} \bigg) \frac{\bar{F}_{\bar{\varphi}}(\bar{\varphi})}{\bar{F}(\bar{\varphi})} \bigg]\,,
\end{equation}
where we have defined $\bar{F}_{\bar{\varphi}}(\bar{\varphi}) \equiv d\bar{F}/d\varphi |_{\bar{\varphi}}$. It was shown in~\cite{Garoffolo:2019mna} that, for this set of theories, the background luminosity distance inferred through the measurement of a GW event is itself modified as
\begin{equation}\label{eq:lumdist_background}
    \bar{D}(\bchi) = \sqrt{\bar{F}(\bar{\varphi})} \, \bar{D}^{\rm EM}(\bchi) = \sqrt{\bar{F}} \, (1+\bar{z}) \int_0^{\bar{z}} dz' \frac{a}{\H (z')}\,,
\end{equation}
where $\bar{D}^{\rm EM}$ can be thought of as the corresponding background luminosity distance that would be probed with an electromagnetic source in the same location, such as a supernova.\footnote{Note that these distances are background quantities, and as such, they cannot actually be measured directly. The correct way to interpret them is as distances that would be measured through GW or SN i) in the absence of large scale structures and ii) if the measurement uncertainties were null.} We can use this relation to calculate explicit expressions for the first and second derivative of the background luminosity distance w.r.t. the background redshift in Eq.~\eqref{eq:counts_general}. For the first derivative we have
\begin{equation}\label{eq:lumdist_background_firstderiv}
    \frac{d\bar{D}}{d\bar{z}} = \sqrt{\bar{F}}\bchi \bigg[1+ \frac{1}{\hchi} - \frac{\dot{\bar{F}}}{2\bar{F}\H}\bigg] = a D \bigg[1+ \frac{1}{\hchi} - \frac{\alpha_M}{2}\bigg]\,,
\end{equation}
where in the last step we have used the definition $\alpha_M \equiv \dot{\bar{F}}/(\bar{F}\H)$. For the subset of theories that we are considering in this study, this function is indeed the same $\alpha_M$ parameter that was first proposed in~\cite{Bellini:2014fua} as part of a minimal parametrization of Horndeski theories. This quantity is a measurement of the running of the Planck mass, more properly defined as $\alpha_M \equiv \H^{-1} d\log M_P^2/d\eta$. This definition in general implies that $\alpha_M$ gets contributions from different operators in the most general Horndeski lagrangian. However, considering Eq.~\eqref{eq:action}, the only surviving function that still contributes to $\alpha_M$ in our case is $F(\varphi)$, so that our definition matches the standard one for the theories explored in this work. In Eq.~\eqref{eq:lumdist_background_firstderiv} we have introduced $\alpha_M$ for two reasons: first, it will lighten significantly the notation for the calculations that follow; additionally, 
it is the function most commonly used in the literature to capture deviations from \lcdm{} through GW at the background level, as it modifies the GW propagation in a simple way, introducing an additional friction term in the propagation equation (\cite{Amendola:2017ovw}).
However, this does \textit{not} mean that from here on we will fully adopt the parametrization of~\cite{Bellini:2014fua}, and indeed, we will clarify this more in detail in the next section. For the time being, the introduction of $\alpha_M$ should be regarded primarily as a convenient way to simplify the notation.

Recalling the definition of $\gamma$~\eqref{eq:gamma_def}, Eq.~\eqref{eq:lumdist_background_firstderiv} leads to
\begin{equation}\label{eq:gamma_st}
    \gamma = \bigg[1+ \frac{1}{\hchi} - \frac{\alpha_M}{2}\bigg]^{-1}\,.
\end{equation}
For the second derivative of the background luminosity distance, we obtain
\begin{equation}\label{eq:lumdist_background_secondderiv}
    \frac{d^2 \bar{D}}{d\bar{z}^2} = a^2 \bar{D} \bigg[ \frac{1}{\hchi} \bigg(1 + \frac{\dot{\H}}{\H^2}\bigg) - \frac{\alpha_M}{2} \bigg( 1+ \frac{2}{\hchi}\bigg) + \frac{\alpha_M^2}{4} + \frac{\dot{\alpha}_M}{2\H}\bigg]\,.
\end{equation}

The last quantity we need to make explicit is the total perturbation of the luminosity distance $\Delta D$. This was calculated in full in the \lcdm{} case in~\cite{Sasaki:1987, Hui:2005nm, Bertacca:2017vod}, and subsequently extended to the class of ST theories considered in this work by~\cite{Garoffolo:2019mna} (hereafter, G20)
\begin{equation}\begin{split}\label{eq:lumdist_perturbations_chi}
    \frac{\Delta D^{\rm G20}(\bchi, \hat{n})}{\bar{D}(\bchi)} &= - \frac{1}{2}\int_0^{\bchi{}} d\chi' \frac{\bchi{}-\chi'}{\bchi{}\chi'} \nabla^2_{\Omega}(\Phi + \Psi) - \Phi + \frac{1}{\bchi}\int_0^{\bchi}d\chi' (\Phi + \Psi) + \\
    &+ \bigg(1 - \frac{1}{\hchi} + \frac{\alpha_M}{2}\bigg) \bigg[\bm{v}\cdot\hat{n} - \Psi - \int_0^{\bchi{}} d\chi' (\dot{\Phi}+\dot{\Psi})\bigg] + \frac{\alpha_M}{2}\H \bigg(\frac{\delta \varphi}{\dot{\varphi}}\bigg)\,,
\end{split}
\end{equation}
where the modifications due to the scalar field manifest explicitly in the second line.

One important consideration about Eq.~\eqref{eq:lumdist_perturbations_chi} is that $\Delta D^{\rm G20} \equiv D[z(\chi), \hat{n}] - \bar{D}[\bz(\bchi)]$ was derived under the condition $1 + z(\chi) = 1 + \bz(\bchi)$\footnote{This can be appreciated in Eq.(5.25) of~\cite{Garoffolo:2019mna}, which also corresponds to Eqs.(18)-(20) in~\cite{Bertacca:2017vod}. A similar condition was also imposed in~\cite{Hui:2005nm}, below Eq.(C17), and in~\cite{Sasaki:1987}, below Eq.(4.7).}. This was done in the spirit of working primarily in terms of observable quantities, where the only relevant redshift is $z$, the observed one.
When observed, this redshift can be interpreted either as the perturbed redshift of a wave propagating along a perturbed geodesics, or the unperturbed redshift of an unperturbed wave originally emitted at redshift $z$. In simpler words, $\Delta D^{\rm G20}$ can be expressed in terms of redshift as $\Delta D^{\rm G20}= D(z, \hat{n}) - \bar{D}(z)$, where $\bD(z)$ is the true distance of a source at true redshift $z$. Therefore, $\Delta D^{\rm G20}$ accounts for the full perturbations of the past lightcone, as can be seen combining Eqs.(5.39) and (5.63) in~\cite{Garoffolo:2019mna},  and to connect it to the $\Delta D$ of Eq.~\eqref{eq:delta_D} required for the number count, it is sufficient to perform a further expansion in terms of the background redshift $\bz$, so that
\begin{equation}\label{eq:redshift_expansion}
    \frac{\Delta D(\bchi, \hat{n})}{\bD(\bchi)} = \frac{\Delta D^{\rm G20}(\bchi, \hat{n})}{\bD(\bchi)} + \frac{1}{\bD(\bchi)} \frac{d\bD}{d z}\bigg|_{z=\bz} \Delta z = \frac{\Delta D^{\rm G20}(\bchi, \hat{n})}{\bD(\bchi)} - \frac{1}{\gamma} \Delta \ln a \, .
\end{equation}
In the last equation, we have used the definition of $\gamma$ Eq.~\eqref{eq:gamma_def}, while $\Delta z$ is the difference between the perturbed and unperturbed redshift of the source, and $\Delta \ln a$ is the corresponding perturbation of the scale factor, that can be expressed in Newtonian gauge as (see~\cite{Garoffolo:2019mna})
\begin{equation}\label{eq:scale_factor}
    \Delta \ln a = \sprod{v}{\hat{n}} - \Psi - \int_0^{\bchi} d\chi' \big(\dot{\Phi}+\dot{\Psi}\big) \, .
\end{equation}
Inserting Eqs.~\eqref{eq:gamma_st},~\eqref{eq:lumdist_perturbations_chi} and~\eqref{eq:scale_factor} in Eq.~\eqref{eq:redshift_expansion}, we finally obtain
\begin{equation}\begin{split}\label{eq:lumdist_perturbations}
    \frac{\Delta D(\bchi, \hat{n})}{\bar{D}(\bchi)} &= - \frac{1}{2}\int_0^{\bchi{}} d\chi' \frac{\bchi{}-\chi'}{\bchi{}\chi'} \nabla^2_{\Omega}(\Phi + \Psi) - \Phi + \frac{1}{\bchi}\int_0^{\bchi}d\chi' (\Phi + \Psi) + \\
    &+ 2 \bm{v}\cdot\hat{n} - 2 \Psi - 2 \int_0^{\bchi{}} d\chi' (\dot{\Phi}+\dot{\Psi}) + \frac{\alpha_M}{2}\H \bigg(\frac{\delta \varphi}{\dot{\varphi}}\bigg)\, .
\end{split}
\end{equation}
Several coefficients have simplified, and now $\alpha_M$ enters only in the coefficient for  the scalar field contribution. 

We compute the derivative of $\Delta D/\bD$ along the null geodesics as
\begin{equation}\begin{split}\label{eq:lumdist_pert_geodesic_deriv}
    \frac{d}{d\bchi} \bigg(\frac{\Delta D}{\bar{D}}\bigg) &=  2 \big(\H{} - \partial_{\bchi} \big ) \vpar - \dot{\Phi} + \bigg(\frac{1}{\bchi} -\partial_{\bchi}\bigg) \Phi + \frac{\Psi}{\bchi} - \frac{1}{2\bchi^2} \int_0^{\bchi} d\chi' \big(2 + \nabla^2_{\Omega}\big) (\Phi + \Psi) \\
    &  - \bigg(\frac{\dot{\alpha}_M}{2} + \frac{\alpha_M}{2}\dot{\H}\bigg) \frac{\delta \varphi}{\dot{\varphi}} - \frac{\alpha_M}{2}\H \frac{d}{d\bar{\eta}}\bigg(\frac{\delta \varphi}{\dot{\varphi}}\bigg)\,.
    \end{split}
\end{equation}
Finally, we substitute Eqs.~\eqref{eq:gamma_st},~\eqref{eq:lumdist_background_secondderiv} and~\eqref{eq:lumdist_pert_geodesic_deriv} into Eq.~\eqref{eq:counts_general} to obtain
\begin{equation}\begin{split}\label{eq:number_count_final}
    \hat{\Delta} (D, \hat{n}) = & \,\,\, \delta_{\rm GW}^{\rm N} + \int_0^{\bchi} d\chi' \bigg[\bigg(\frac{\beta(\bchi)-1}{2}\bigg) \frac{\bchi - \chi'}{\bchi\chi'} + \frac{\gamma(\bchi)}{2\H\bchi^2}\bigg] \nabla^2_{\Omega} (\Phi + \Psi) - 2 \frac{\gamma}{\H} \,  \partial_{\bchi} (\bm{v}\cdot\hat{n}) + \\[8pt]
    &  - \bigg[1 + 2 \gamma  + 2 \beta \bigg] \bm{v}\cdot\hat{n} + \bigg[\frac{1-\beta}{\bchi} + \frac{\gamma}{\H\bchi^2}\bigg] \int_0^{\bchi} d\chi' (\Phi + \Psi) + 2 \big(\beta + 1\big) \int_0^{\bchi} d\chi'(\dot{\Phi} + \dot{\Psi}) \\[8pt]
    & + \bigg[\beta - 1 - \frac{\gamma}{\bchi\H}\bigg] \Phi + \frac{\gamma}{\H} \partial_{\bchi} \Phi + \frac{\gamma}{\H} \dot{\Phi} + \bigg[1-\frac{\gamma}{\H\bchi} + 2 \big(\beta + 1\big)\bigg] \Psi +\\[8pt]
    &+ \gamma \frac{\alpha_M}{2} \frac{d}{d\bar{\eta}}\bigg(\frac{\delta \varphi}{\dot{\varphi}}\bigg) + \bigg[\gamma \frac{\dot{\alpha_M}}{2} + \frac{\alpha_M}{2} \bigg(\gamma \frac{\dot{\H}}{\H}-\beta - 1\bigg) \bigg] \frac{\delta \varphi}{\dot{\varphi}}\,,
\end{split}
\end{equation}    
where we have further defined
\begin{equation}\begin{split}
    \beta &\equiv \gamma \bigg(\frac{2}{\bchi{} \H{}} + \frac{\dot{\H{}}}{\H{}^2} - \frac{\gamma}{a^2 \bar{D}} \frac{d^2\bar{D}}{d\bar{z}^2} -1 - \frac{\partial\ln \bar{n}}{\partial\ln a} \bigg) \\
    &= \gamma \bigg\{\frac{2}{\bchi{} \H{}} + \frac{\dot{\H{}}}{\H{}^2} - \gamma \bigg[ \frac{1}{\hchi} \bigg(1 + \frac{\dot{\H}}{\H^2}\bigg) - \frac{\alpha_M}{2} \bigg( 1+ \frac{2}{\hchi}\bigg) + \frac{\alpha_M^2}{4} + \frac{\dot{\alpha}_M}{2\H}\bigg] -1 - \frac{\partial\ln \bar{n}}{\partial\ln a}\bigg\}\,.
\end{split}
\end{equation}
Eq.~\eqref{eq:number_count_final} represents the final theoretical expression, at linear order, for the number count of gravitational waves in the set of scalar-tensor theories of gravity considered. We can recognise a number of perturbative terms contributing to the number count, analogous to the ones affecting galaxies (see e.g.~\cite{Bonvin2011}). All of them, except from the GW clustering $\delta_{\rm GW}^N$, that constitutes the core of our signal, are induced by relativistic effects responsible for the GW perturbed propagation. In order, these terms are: the density contrast of GW events in Newtonian gauge, the lensing term, the luminosity distance space distortions (DSD) term - analogous to the redshift space distortions of galaxies - the Doppler term, the Shapiro time delay term, the ISW term, four local potential terms containing the metric perturbations $\Psi$ and $\Phi$ and their derivatives evaluated at the source position, and two additional terms, that vanish in the \lcdm{} limit, which are the relativistic effects linked to variations in the scalar field. We stress that these last two terms are not the only explicit contribution to the GW number count introduced by the scalar field, as the coefficient of all terms contain the functions $\gamma$ and $\beta$, which in turn explicitly contain $\alpha_M$. Moreover, each of the terms will be implicitly affected by the modifications of gravity, as the perturbations grow and evolve differently in different theories. As such, the GW number count has the potential to become a valuable probe to test gravity models on cosmological scales. We will investigate this potential in the next sections.

We note that our calculation for the number count does not immediately apply to other EM sources living in luminosity distance space, such as supernovae. Indeed, the propagation of photons is affected differently by modifications of gravity. At the background level, the luminosity distance of an EM source is defined in Eq.~\eqref{eq:lumdist_background}, and this will affect the following Eqs.~\eqref{eq:lumdist_background_firstderiv},~\eqref{eq:gamma_st} and~\eqref{eq:lumdist_background_secondderiv}, where the $\alpha_M$ contribution vanishes. Eq.~\eqref{eq:lumdist_perturbations} is also different in the case of photons: here $\alpha_M=0$ implies that there is no perturbative term due to the fluctuations of the scalar field $\delta \varphi$. The net effect of these differences on the final result however is quite easy to work out. Indeed, the number count of EM sources in ST theories will just be Eq.~\eqref{eq:number_count_final}, with the coefficients of all remaining terms taken in the \lcdm{} limit, i.e. it will be
\begin{equation}
    \gamma^{\rm SN} = \bigg[1+ \frac{1}{\hchi}\bigg]^{-1} \quad \text{and} \qquad \beta^{\rm SN} = \gamma \bigg\{\frac{2}{\bchi{} \H{}} + \frac{\H{}'}{\H{}^2} - \frac{\gamma}{\hchi} \bigg(1 + \frac{\dot{\H}}{\H^2}\bigg) -1 - \frac{\partial\ln \bar{n}}{\partial\ln a}\bigg\}\,. 
\end{equation}

These differences open the interesting possibility of adding constraining power over MG models not only by combining the GW signal with galaxies, but also with other EM sources living in luminosity distance space, similarly to what was done in~\cite{Garoffolo:2020vtd} to isolate the scalar field fluctuations.

Let us conclude this part, noticing that the \lcdm{} limit of Eq.~\eqref{eq:number_count_final} does not match the results of~\cite{Fonseca:2023uay} (Hereafter, F23) - note that their definition of $\beta$ is such that $\beta^{\rm F23} = \beta -1$. The discrepancy is due to the sign difference pointed out in Eq.~\eqref{eq:counts_general}, propagated to the final result, and affects all terms involving $\beta$, i.e. lensing, Doppler, time delay, ISW and two potential terms. Though, as we will show, most of these terms are negligible corrections to the number count, the differences induced on the lensing term are important. We comment more on this in Sec.~\ref{sec:synergies}.

\subsection{Magnification and evolution biases}
\label{subsec:biases}

The observation of any object in the sky is limited by a detector sensitivity. For both supernovae and gravitational waves, this takes the form of a minimum flux, below which the object goes undetected. As flux is linked to the luminosity distance of the source, a variation in the observed luminosity distance induced by perturbations can change the observed flux, pushing sources above or below the detection limit, and in turn affecting the number count. This effect is widely known even in the case of galaxy surveys, and goes by the name of magnification bias. The magnification bias for the luminosity distance number count was derived for a \lcdm{} universe in~\cite{Fonseca:2023uay}. Here, we retrace their derivation, highlighting, if any, the additional contributions coming from the presence of the scalar field. 

For SN, the flux limit of the survey is translated to an upper limit on their apparent magnitude $m$, which in turn is connected to the luminosity distance of the source through
\begin{equation}\label{eq:apparent_magnitude}
    M = m - 5\log_{10} (D/{\rm Mpc}) + 25 \, .
\end{equation}
Being SN standard candles, any perturbation in the luminosity distance is translated into a corresponding perturbation to the apparent magnitude
\begin{equation}
    m(D) = m(\bD) + \frac{\partial m}{\partial D}\bigg|_{\bD} \Delta D  = m(\bD) + \frac{5}{\ln 10} \frac{\Delta D}{\bD} \equiv m(\bD) + \Delta m (\bD) \, .
\end{equation}
This in turn implies a correction to the  SN number density, i.e. Eq~\eqref{eq:density_perturbations_def} in App.~\ref{app:density_perturbations} becomes
\begin{equation}
    \frac{\rho(D, \hat{n}, m) - \bar{\rho}(D, m)}{\bar{\rho}(D, m)} = \delta^N_{\rm GW} - \frac{1}{\bar{\rho}}\frac{d\bar{\rho}}{d\bar{D}}\bigg|_{D=\bar{D}} \Delta{D} - \frac{1}{\bar{\rho}}\frac{d\bar{\rho}}{dm}\bigg|_{m(\bD)} \Delta m (\bD)  \, .
\end{equation}
Remembering that $\bar{\rho} = a^{-3} \bar{n}$, with $\bar{n}$ the comoving density of sources, the net perturbation to the number density can be written as
\begin{equation}
    \hat{\Delta}^{\rm SN}(\hat{n}, D, m) = \hat{\Delta}^{\rm SN}(\hat{n}, D) - 5 \, \frac{\partial \log_{10} \bar{n}}{\partial m}\bigg|_{m(\bD)} \frac{\Delta D}{\bD} \, .
\end{equation}
Any MG effect will enter explicitly in the above expression through the luminosity distance perturbations Eq.~\eqref{eq:lumdist_perturbations}. 

Gravitational waves have also a detection threshold, which is typically chosen to be a minimum value for the signal-to-noise of the GW in the detector. This SNR $\Theta$ is defined as
\begin{equation}\label{eq:snr_def}
    \Theta^2 \equiv 4 \int_0^\infty \frac{|h(f)|^2}{S_n(f)} df \, ,
\end{equation}
where $S_n(f)$ is the noise spectral density of the specific detector, and $h(f)$ is the GW strain in frequency domain, which in general reads~\cite{Robson:2018ifk}
\begin{equation}
    h(f) = \mathcal{A}(f) \bigg( F_{+}(f) \frac{1+\cos^2 \iota}{2} + i F_{\times}(f) \cos \iota \bigg) e^{i\psi(f)} \, .
\end{equation}
Here, $\mathcal{A}(f)$ and $\psi(f)$ are respectively the amplitude and phase of the GW wave, $\iota$ is the inclination of the orbital plane of the binary with respect to the line of sight and $F_{+}$ and $F_{\times}$ are the frequency-dependent pattern functions that model the detector response to each polarization of the wave. Then, the signal spectral power becomes
\begin{equation}
    |h(f)|^2 = \mathcal{A}^2(f) \bigg[ F_{+}^2(f) \bigg(\frac{1+\cos^2\iota}{2}\bigg)^2 + F^2_{\times}(f) \cos^2\iota\bigg] \, ,
\end{equation}
i.e., for the SNR we can write
\begin{equation}\label{eq:snr_final}
    \Theta^2 = \frac{4}{D^2} \int_0^\infty  df \frac{\mathcal{Q}^2(f)}{S_n(f)} \bigg[ F_{+}^2(f) \bigg(\frac{1+\cos^2\iota}{2}\bigg)^2 + F^2_{\times}(f) \cos^2\iota\bigg] \propto \frac{1}{D^2} \, ,
\end{equation}
where we have used that, at any post-newtonian order, the amplitude of the wave can always be written as $\mathcal{A}(f) = \mathcal{Q}(f)/D$, with $\mathcal{Q}(f)$ a function including in its definition the intrinsic parameters of the binary. From Eq.~\eqref{eq:snr_final}, we now appreciate that a perturbation on the luminosity distance induces a fluctuation in the GW SNR that can be written as
\begin{equation}
    \Theta(D) = \Theta(\bar{D}) + \frac{\partial \Theta}{\partial D}\bigg|_{\bD} \Delta D \equiv \Theta(\bar{D}) + \Delta \Theta (\bD) \, ,
\end{equation}
from which we obtain
\begin{equation}
    \frac{\Delta \Theta(\bD)}{\Theta (\bD)} = -\frac{\Delta D}{\bD} \, .
\end{equation}
In total analogy with SN surveys, the corresponding perturbation produced on $\hat{\Delta}^{\rm GW}(D, \hat{n})$ then reads
\begin{equation}
    \hat{\Delta}^{\rm GW} (D, \hat{n}, \Theta) = \hat{\Delta}^{\rm GW}(D, \hat{n}) - \frac{\partial \ln \bar{n}}{\partial \ln \Theta}\bigg|_{\Theta(\bD)} \frac{\Delta \Theta(\bD)}{\Theta(\bD)} = \hat{\Delta}^{\rm GW}(D, \hat{n}) + \frac{\partial \ln \bar{n}}{\partial \ln \Theta}\bigg|_{\Theta(\bD)} \frac{\Delta D}{\bD} \, .
\end{equation}
This expression carries the imprint of deviations from \lcdm{} at two levels: first, through the luminosity distance perturbations Eq.~\eqref{eq:lumdist_perturbations}; then, because of the implicit dependence of $\Theta(\bD)$ on the non-minimal coupling, through Eq.~\eqref{eq:lumdist_background}, which can in turn impact the SNR evolution of the comoving density of sources, $\partial \ln n / \partial \ln \Theta$. 

The magnification bias discussed in this section affects the luminosity distance number count in addition to the evolution bias $\partial \ln \bar{n} / \partial \ln a$ presented in Sec.~\ref{sec:reference_frames}. The proper modeling of both quantities for GW requires the accurate characterization of the comoving density of events, which in turns needs to rely heavily on astrophysical models for the redshifts evolution of the binary rates.
This task is complex and delicate, and we do not address it in the present study, though we stress its investigation is an important ingredient of any complete cosmological forecast analysis. See~\cite{Libanore:2021jqv, Zazzera:2023kjg} for some recent work in this direction. For the rest of the paper, we will simply set
\begin{equation}
    \frac{\partial \ln \bar{n}}{\partial \ln a} = \frac{\partial \log_{10} \bar{n}}{\partial m} = \frac{\partial \ln \bar{n}}{\partial \ln \Theta} = 0 \, .
\end{equation}

\section{Tomographic observables in Scalar-Tensor theories}
\label{sec:correlations}

The number count $\hat{\Delta}(D, \hat{n})$ is not yet the actual observable. This can be obtained by integrating Eq.~\eqref{eq:number_count_final} over an observed luminosity distance window function, so that we get
\begin{equation}\label{eq:observable_number_count}
    \Delta^{\rm obs} (D^i, \hat{n}) = \int_0^{\infty} dD \, W^i(D) \hat{\Delta} (D, \hat{n}) = \int_0^{\infty} d\bD \, W^i(\bD) \hat{\Delta} (\bD, \hat{n})\,,
\end{equation}
where we are taking a tomographic approach and $W^i(D)$ becomes then the window function identifying the $i^{th}$ tomographic luminosity distance bin, centered around $D^i$. Binning has the unfortunate effect of partially smoothing out the signal, decreasing its strength, but is necessary as this window accounts for the measurement uncertainty on the observed luminosity distances, which depends on the specific detector considered. The smaller this error, the finer the binning can be, the strongest the observed number count signal. The last equality in Eq.~\eqref{eq:observable_number_count} holds as we are doing a linear calculation and $\hat{\Delta}$ is already a first order perturbation. In what follows, as there is no more room for confusion, we drop the bar to indicate background quantities and always implicitly intend all objects evaluated on the background.

Eq.~\eqref{eq:observable_number_count} is correctly integrated in luminosity distance space, where $\hat{\Delta}$ can be measured. However, we are interested in exploring the potential of this observable as a cosmological probe, by itself and in correlation with other tracers, such as galaxies and SN. To this extent we  re-parametrize this integral in terms of conformal time
\begin{equation}\label{eq:observable_window}
    \Delta^{\rm obs} (D^i, \hat{n}) = \int_0^{\infty} dD \,  W^i(D) \, \hat{\Delta}(D, \hat{n}) = \int_0^{\eta_0} d\eta \, W^i(\eta) \hat{\Delta}(\eta, \hat{n})\,,
\end{equation}
with $D = (\eta_0 - \eta) \sqrt{F} / a$, $\eta_0$ being the conformal time at observation and
\begin{equation}\label{eq:window_def}
   W^i(\eta) \equiv \bigg| \frac{dD}{d\eta} \bigg| \, W^i(D(\eta)) = \frac{\sqrt{F} (\eta_0 - \eta) \H}{a} \bigg[1+ \frac{1}{(\eta_0-\eta) \H} - \frac{\alpha_M}{2}\bigg] \, W^i\big(D(\eta)\big)\,.
\end{equation}
These expressions can then be implemented in an Einstein-Boltzmann solver; in our case, we employ \texttt{EFTCAMB}~\cite{Hu:2013twa, Raveri:2014cka}, an effective field theory extension of the publicly available \texttt{CAMB}~\cite{Lewis:1999bs,Howlett:2012mh}. 

The standard cosmological observables associated with the number count are the 2-point angular correlation functions. To extract these correlations from Eq.~\eqref{eq:observable_window}, we manipulate $\Delta^{\rm obs}$ going to Fourier-space, and expanding each mode in spherical harmonics such that
\begin{equation}\begin{split} \label{eq:spherical_coefficients_def}
    \Delta^{\rm obs} (D^i,\hat{n}) &= \sum_{\ell=0} \sum_{m=-\ell}^{\ell} \bigg[ 4\pi i^{\ell} \int \frac{d^3 k} {(2\pi)^{3/2}} Y^*_{\ell m}(\hat{k}) \int_0^{\eta_0} d\eta \, W^i(\eta) \, \hat{\Delta}(\eta, {\bm k}) \, j_{\ell}(k\chi) \bigg] Y_{\ell m}(\hat{n}) \\
    & \equiv \sum_{\ell=0} \sum_{m=-\ell}^{\ell} \Delta_{\ell m}^{\rm obs}(D^i) \, Y_{\ell m}(\hat{n})\,.
\end{split}
\end{equation}

The coefficients $\Delta^{\rm obs}_{\ell m} (D^i)$ are given by the term in squared parenthesis.  The 2-point correlation functions are then defined in the standard way as
\begin{equation}
    \langle \Delta^{\rm obs}_{\ell m} (D^i) \big(\Delta^{\rm obs}_{\ell' m'}(D^j)\big)^*\rangle \equiv \delta_{\ell \ell'} \delta_{m m'} \, C^{ij}_\ell\,,
\end{equation}
where the bracket on the LHS indicates an average over large volumes. Computing the average and exploiting the normalization properties of the spherical harmonics, it follows easily that
\begin{equation}\label{eq:cross_correlations}
    C^{ij}_{\ell} = 4\pi \int d \log k \int_0^{\eta_0} d\eta \, W^i(\eta) \, j_\ell(k\chi)  \int_0^{\eta_0} d\eta' \, W^j(\eta') \, j_\ell(k, \chi') \,\, P_{\hat{\Delta}}(k,\eta, \eta') \,,
\end{equation}
where $P_{\hat{\delta}}(k, \eta, \eta')$ is the a-dimensional power spectrum defined as
\begin{equation}
    \langle \hat{\Delta}(\eta, \bm{k}) \hat{\Delta}^* (\eta', \bm{k'}) \rangle = \frac{2\pi^2}{k^3} P_{\hat{\Delta}}(k, \eta, \eta') \, \delta_D^{(3)}(\bm{k}-\bm{k'})\,,
\end{equation}
connected to the primordial power spectrum $P_P(k) \equiv \mathcal{A_s} \big(k/k_p\big)^{n_s-1}$ through
\begin{equation}
    P_{\hat{\Delta}}(k, \eta, \eta') = \Delta^T (k, \eta) \Delta^T (k, \eta') P_P(k)\,,
\end{equation}
where $\Delta^T$ is the transfer function encoding the time evolution of $\hat{\Delta}$ from the primordial inflationary perturbation.

\subsection{Scalar-Tensor models in the EFT formalism}
\label{STmodels}
We work with \texttt{EFTCAMB}~\cite{Hu:2013twa, Raveri:2014cka} (a patch to the publicly available Einstein-Boltzmann solver \texttt{CAMB}~\cite{Lewis:1999bs, Howlett:2012mh}), which uses the framework of effective field theory of dark energy (EFTofDE) ~\cite{Cheung:2007st, Creminelli:2008wc, Park:2010cw, Bloomfield:2011np, Gubitosi:2012hu, Piazza:2013coa, Gleyzes:2013ooa} to explore ST theories.  We adopt the convention for the EFTofDE action defined in~\cite{Hu:2013twa,Raveri:2014cka}, where background and linear dynamics for our action~(\ref{eq:action}) are equivalently described by the following quadratic action:
\begin{equation}\begin{split}\label{EFT_action}
\mathcal{S} = \int d^4x \sqrt{-g}  \bigg\{ &\frac{m_0^2}{2} (1+\Omega(\tau))R + \Lambda(\tau) - c(\tau)\,a^2\delta g^{00} + \gamma_1(\tau)\frac{m_0^2H_0^2}{2} \left(a^2 \delta g^{00} \right)^2\\
 &- \gamma_2(\tau)\frac{m_0^2H_0}{2} \, a^2\delta g^{00}\,\delta {K}{^\mu_\mu} +	\ldots \bigg\}+S_{m} [g_{\mu \nu}, \chi_m ]\,,
\end{split}
\end{equation}
where  we have adopted the notation used in \texttt{EFTCAMB}~\cite{Hu:2013twa,Raveri:2014cka} and  $m_0=(8\pi G)^{-1}$. The corresponding running Planck mass is $M^2_P=m_0^2(1+\Omega)$, from which it follows that 
\begin{equation}
    \alpha_M(\eta) = \frac{\dot{\Omega}(\eta)}{\H(\eta)\left[1+\Omega(\eta)\right]}\,,
\end{equation}
which is consistent with the fact that the mapping of Eq.~\eqref{eq:action} into Eq.~\eqref{EFT_action} gives $F(\eta) = m_0/2 \big(1 + \Omega(\eta)\big)$.

The free functions of time multiplying the different operators, $\{\Omega, c,\Lambda, \gamma_1, \gamma_2\}$ are dubbed \emph{EFT functions}. The first three impact both the background and linear dynamics, while the $\gamma$s affect only perturbations. We work in the so-called designer approach, where we make a choice for the expansion history and $\Omega$, and rely on the Friedmann equations to correspondingly fix $\Lambda$ and $c$.  Consequently, the  full phenomenology of action Eq.~\eqref{eq:action} on linear scales can be reproduced by varying four free functions of time: $\{\Omega, \Lambda, \gamma_1,\gamma_2\}$. 
 
We explore different ST models belonging to Eq.~\eqref{eq:action}, via a parametrization of the  expansion history and $\{\Omega(\eta),\gamma_1(\eta),\gamma_2(\eta)\}$. For the background, we parametrize the equation of state of dark energy following  ~\cite{Chevallier:2000qy,Linder:2002et}
\begin{equation}\label{eq:dark_energy_eos}
    w_{\rm DE}(a) = w_0 + w_a(1-a) \, , \qquad \text{leading to} \qquad \rho_{\rm DE} (a) = \rho^0_{\rm DE} a^{-3 (1+w_0+w_a)}e^{-3w_a(1-a)}\,.
\end{equation}
Here, $\rho_{\rm DE}$ is the background dark energy density, and $\rho_{\rm DE}^0$ its value today.
For the EFT function, we choose:
\begin{equation}\label{eq:par_EFTfunction}
 \Omega(a) \equiv \Omega_0 \, \frac{\rho_{\rm DE} }{\rho^0_{\rm DE}}\,, \qquad  \gamma_1(a) \equiv \gamma_{1,0} \, \frac{ \rho_{\rm DE}}{\rho^0_{\rm DE}}\,, \qquad  \gamma_2(a) \equiv \gamma_{2,0} \, \frac{ \rho_{\rm DE}}{\rho^0_{\rm DE}}\,.
\end{equation}

With this set of definitions, any MG model is completely specified with the choice of five parameters: $w_0$, $w_a$, $\Omega_0$, $\gamma_{1,0}$ and $\gamma_{2,0}$. Additionally, we must consider the six additional parameters that determine the standard background cosmology. These are: the present value of the Hubble parameter $H_0$, the density contrast of matter $\Omega_{m,0}$, baryons $\Omega_{b,0}$ and curvature $\Omega_{K,0}$ today, the primordial power spectrum spectral index $n_s$ and the amplitude $\sigma_8$ of the linear matter power spectrum at $8 {\rm Mpc}/h$. We keep these parameters fixed at the best fit of Planck (2018)~\cite{Planck:2018vyg}.

\begin{comment}
\begin{table}
    \renewcommand{\arraystretch}{1.2}
    \centering
    \begin{tabular}{|l|c|c|c|c|c|c|}
        \hline
        Model & $w_0$ & $w_a$ & $\Omega_0$ & $\gamma_1^0$ & $\gamma_{2,0}$ & $B_0$ \\
        \hline
        \hline
        \lcdm{} & -1.0 & 0.0 & 0.0 & 0.0 & 0.0 & - \\
        CPL & -0.9 & -0.05 & 0.0 & 0.0 & 0.0 & - \\
        GBD & -0.946 & -0.098 & 0.018 & 0.0 & 0.0 & - \\
        KGB & -0.94 & -0.31 & 0.047 & 4.4 & -0.23 & - \\
        f(R) & -1.0 & 0.0 & - & - & - & $10^{-5}$ \\
        \hline
    \end{tabular}
    \caption{Fiducial parameter values for the cosmological models considered in this work. The background parameters $h$, $\Omega_{m,0}$, $\Omega_{b,0}$, $n_s$ and $\sigma_8$ were kept fixed at Planck18~\cite{Planck:2018vyg} best fit values.}
    \label{tab:models_fiducial}
\end{table}
\end{comment}

\begin{table}
    \renewcommand{\arraystretch}{1.3}
    \centering
    \begin{tabular}{l|c|c|c|c|c|c}
        \hline
        \hline
        Model & $w_0$ & $w_a$ & $\Omega_0$ & $\gamma_1^0$ & $\gamma_{2,0}$ & $B_0$ \\
        \hline
        \lcdm{} & -1.0 & 0.0 & 0.0 & 0.0 & 0.0 & - \\
        $w$CDM & -0.9 & -0.05 & 0.0 & 0.0 & 0.0 & - \\
        GBD & -0.946 & -0.098 & 0.018 & 0.0 & 0.0 & - \\
        KGB & -0.94 & -0.31 & 0.047 & 4.4 & -0.23 & - \\
        f(R) & -1.0 & 0.0 & - & - & - & $10^{-5}$ \\
        \hline
        \hline
    \end{tabular}
    \caption{Fiducial parameter values for the cosmological models considered in this work. The background parameters $h$, $\Omega_{m,0}$, $\Omega_{b,0}$, $n_s$ and $\sigma_8$ were kept fixed at Planck18~\cite{Planck:2018vyg} best fit values.}
    \label{tab:models_fiducial}
\end{table}

We explore the number count correlations in five different models within action~\eqref{eq:action}, namely: the \lcdm{} cosmological model; a $w$CDM  model, i.e. a model with a non-trivial dark energy equation of state, but all EFT functions set to zero; a Generalized Brans-Dicke model (GBD,~\cite{DeFelice:2010jn}), characterised by non-minimal coupling and a standard kinetic term; a Kinetic Gravity Braiding model (KGB,~\cite{Deffayet:2010qz});  and an $f(R)$ model (\cite{Song:2006ej}) on a \lcdm{} background. For the latter, we adopt the built-in \textit{full mapping} module of \texttt{EFTCAMB}~\cite{Hu:2014oga}, that implements a family of designer $f(R)$ models on \lcdm{} background, labeled by the parameter $B_0$. This parameter is representative of the mass scale of the scalar degree of freedom today, in units of horizon scale, and can be related to $\Omega_0$~\cite{Hu:2014oga}. 
We summarize the fiducial values of the DE and EFT parameters adopted for each model in Tab.~\ref{tab:models_fiducial}. For GBD and KGB we chose as fiducial model the best fit values of~\cite{Frusciante:2018jzw}, while the fiducial for $f(R)$ is chosen to be compatible with the bounds of~\cite{Raveri:2014cka}.

\subsection{GW, SN and galaxies: angular correlations in D-space vs z-space}

In this section and the following, we plot different quantities related to the theoretical correlations of Eq.~\eqref{eq:cross_correlations}. In doing so, we make a number of choices. First, as discussed in Sec.~\ref{subsec:biases} we neglect the evolution and magnification biases, i.e. we fix
\begin{equation}
    \frac{\partial \ln\bar{n}_{\rm [GW,SN]}}{\partial \ln a} = 0 \, , \qquad \frac{\partial \ln\bar{n}_{\rm GW}}{\partial \ln \sigma} = 0 \, , \qquad \text{and} \qquad \frac{\partial \ln\bar{n}_{\rm SN}}{\partial \ln m} = 0\,.
\end{equation}

For the bias factor $b_{\rm gw}$ linking the GW density contrast  to the underlying dark matter density contrast, $\delta_{\rm gw}^S = b_{\rm gw} \delta_m^S$ (and similar for SN), for simplicity, in this work we limit to $b_{\rm gw} = b_{\rm sn} = 1$.
Then, we need to fix the shape of the window function in Eq.~\eqref{eq:window_def}. We choose the adopt the redshift gaussian window function already implemented in \texttt{CAMB}, i.e. we use
\begin{equation}\label{eq:gaussian_window}
    W^i(z) = C \exp \bigg[\frac{(z-z^i)^2}{2(\sigma^i_z)^2}\bigg] \qquad \rightarrow \qquad W^i(\eta) = \bigg|\frac{dz}{d\eta}\bigg| W^i(z)\,,
\end{equation}
with each gaussian bin having center $z^i$ and width $\sigma^i_z$. Since we are concerned with  a theoretical analysis, we can parametrize the window function in terms of the redshift of the source exploiting the background relation $D^i(z^i)$ of Eq.~\eqref{eq:lumdist_background}.
This way, we can study how our expected signal changes as a function of redshift, providing a direct comparison with the number count of galaxies. Moreover, using the exact same window to calculate angular correlations for GW/SN and galaxies will facilitate the comparison of the two. In practice, of course, the GW redshift will not be available, and in a forecast analysis we should use a window parametrized in terms of $D$, as in Eq.~\eqref{eq:window_def}. For the main part of our analysis, we will use $\sigma^i_z =0.2$, unless specified otherwise. We will analyze further the implications of this choice in Sec~\ref{subsec:binning}. 

\begin{figure}
    \centering
    \includegraphics[width=1.\textwidth]{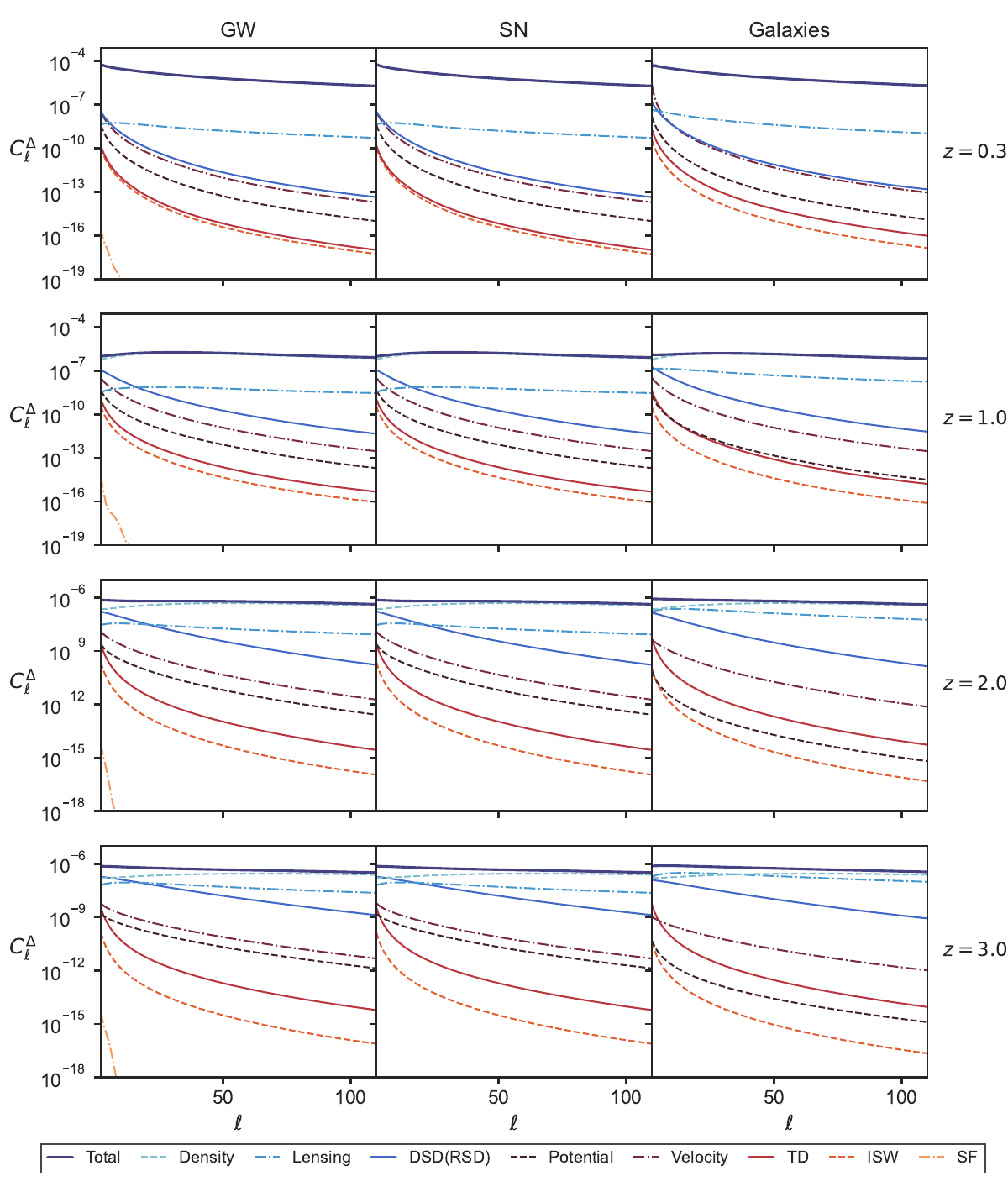}
    \caption{Angular correlations of the number counts of GW (left), SN (center) and galaxies (right) in the fiducial GBD model. Each curve shows the auto-correlations of a specific relativistic effect, i.e. the number count signal obtained switching off all other perturbations. The dark blue solid curve shows instead the total signal. We use GBD to show the possible difference between GW and SN, and the individual contribution of the scalar field clustering. Comparing with the \lcdm{} case, we see no difference in these curves appreciable by eye, except the absence of the scalar field curve in left column panels, thus this plot can also be used as representative of \lcdm{}.}
    \label{fig:correlations}
\end{figure}

In Fig.~\ref{fig:correlations}, we plot the auto-correlations of the number counts signal for GW, SN, and galaxies,  at different redshifts.  The dark  solid line in the top of each plot shows the total signal, while the other  lines  correspond to the auto-correlations of each perturbative component of the signal, i.e. the auto-correlations of the number counts obtained switching off all perturbative terms in Eq.~\eqref{eq:number_count_final} except the one under consideration. We choose to show the results directly in GBD instead of \lcdm{}, because in this model we have a non-vanishing $\alpha_M$, which potentially allows us to show a difference between the number count of GW and SN, and additionally sources the scalar field clustering contribution to the total signal of GW. We find, however, that the deviations from the \lcdm{} curves are small, as we will show in more detail below. So small in fact that , this plot can be considered also representative of the \lcdm{} scenario (with the only difference that there would be no scalar field curve in the left column panels).

From the Figure, one can see that the total signal for GW and SN at low and intermediate redshifts ($z = 0.3-1.0$) is comparable to the galaxy signal, with the galaxy one being only slightly higher  at $\ell < 25$. For these redshifts, the density contrast dominates the signal, except for a small contribution from the lensing and velocity terms at very low multipoles. For all sources, the lensing auto-correlation  becomes more relevant as the redshift increases; this is the case also for the other integrated effects, i.e. the time delay  for which the auto-correlation changes significantly at different redshifts, and the ISW  that grows  mildly in redshift. We notice that, comparing our results with~\cite{Fonseca:2023uay}, in our case the lensing auto-correlation grows way slower in redshift, remaining about $1-2$ orders of magnitude lower than the density contrast correlation even at $z=3.0$. We find that that is imputable to the sign difference we spotted in Eq.~\eqref{eq:counts_general}, propagated on the final amplitude of the lensing convergence term. We caution that this also impacts the total number count correlation in possibly a non-negligible way, as consequently the lensing convergence will carry a higher weight on the total signal. As for the local effects, the Doppler auto-correlation term  decreases with increasing redshifts, consistent with the impact of the peculiar velocity becoming less and less relevant, while the potential terms auto-correlation  remain essentially the same at different redshifts. The DSD correlation of GW and SN, similarly to the galaxy case, increase at increasing redshifts. Finally, we see that the scalar field clustering contributes very little to the GW number counts, being several orders of magnitude lower than the signal from any other perturbation.

\begin{figure}
    \centering
    \includegraphics[width=1.\textwidth]{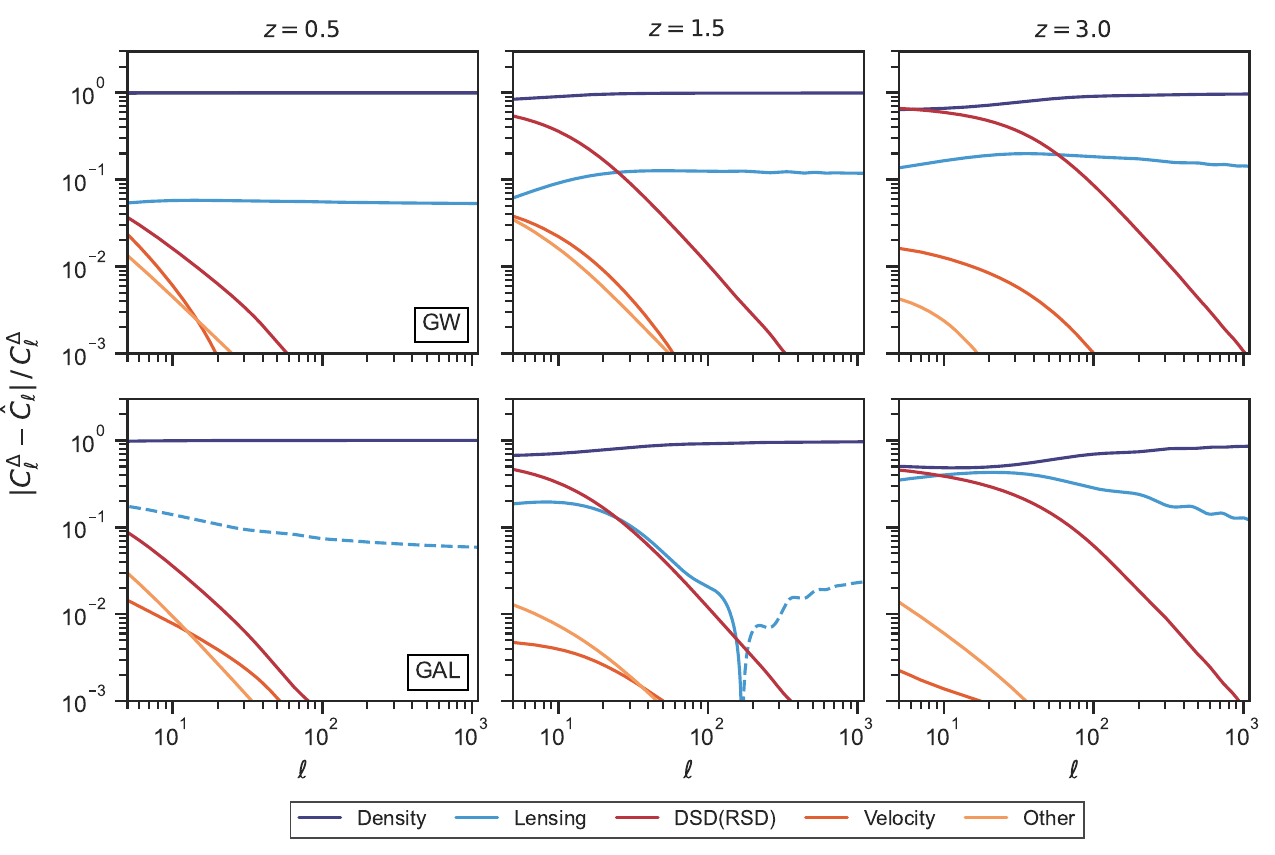}
    \caption{Contribution of the individual perturbation terms to the number count signal in \lcdm, as a fraction of the total signal. The fractional contribution is computed as $|C_{\ell} - \hat{C}_{\ell}|/C_{\ell}$, where $\hat{C}$ is the number count signal obtained including all perturbations effects except the selected one. Lines of different color refer to different perturbation terms (with the yellow lines indicating the contribution of all term except density contrast, lensing, DSD/RSD and Doppler), while the dashed line indicates where the lensing contribution has negative sign.
    The panels in the first row refer to GW, while in the bottom row are the analogous plots for the galaxies. Finally, different panels correspond to different redshifts, as indicated, in the range $[0.5, 3.0]$.}
    \label{fig:MG_contributions}
\end{figure}

Though rich in information, Fig.~\ref{fig:correlations} still makes it difficult to quantify how much the different perturbative terms contribute to the total signal, as it shows only the auto-correlation of the individual perturbations. However, also their cross-correlation with all other perturbations can carry significant weight. We quantify more accurately the contribution of each perturbation in Fig.~\ref{fig:MG_contributions}, where we plot it as a fraction of the total signal, at different redshifts from $0.5$ to $3.0$. Specifically, for each contribution to the number count, we plot the quantity $|C^{\Delta}_{\ell}-\hat{C}_{\ell}|/C^{\Delta}_{\ell}$, where $C^{\Delta}_{\ell}$ is the total auto-correlation of the number count, while $\hat{C}_{\ell}$ is the same auto-correlation, but obtained removing from the number count the perturbation under analysis. For example, in the case of  DSD $\hat{C}_{\ell}$ would be the correlation of $[\hat{\Delta}+ \gamma/\H \, \partial_{\chi}\, (\bm{v}\cdot\hat{n})]$.  We group together the  scalar field,  ISW,  time delay and potentials terms since their contributions are significantly smaller than the others, as can be anticipated from Fig.~\ref{fig:correlations}.

We explore the fractional contributions in \lcdm{}, and compare the  GW  and galaxies case. 
We  can see that at low redshifts, the number count signal is strongly dominated by the density contrast. At higher redshifts, the other components start to become important, especially on larger angular scales. In the case of GW, at $\ell \geq 100$ the second most important component is always lensing, which contributes about $6\%$ of the signal at low redshifts, growing up to $10-20 \%$ at $z=1.5$ and $z=3.0$. The other most important contributors are the DSD, which however are only relevant on larger scales ($\ell \leq 100$). At low $z$, the DSD contribution remains below $\sim 5 \%$, while it weights up to $50-70\%$ on the total signal at higher $z$ and low multipoles. Finally, the Doppler term always remains subdominant with respect to lensing and DSD, contributing no more than $3\%$ of the signal, at all redshifts. Similarly, all other perturbations jointly only contribute, at best, a few $\%$ of the signal, and only for $\ell \leq 20$. We note that Fig.~\ref{fig:MG_contributions}, aside from giving a visualization of the dominant components of the number count, can also be used to quantify the error committed in neglecting some of the less relevant perturbative terms in a cosmological analysis. Neglecting terms can sometimes be useful, as it can speed up the numerical calculation of the angular correlations, but it comes at the price of loosing accuracy. Whether or not this is acceptable, needs to be decided on a case-by-case basis, depending on the specific GW detector targeted, which also determines the noise associated to the correlation, as we will discuss in Sec.~\ref{sec:observations}. Generally speaking, we conclude from Fig.~\ref{fig:MG_contributions} that in any accurate analysis, the density, DSD and lensing perturbations should all always be included. The other perturbative terms can sometimes be safely neglected, depending on the required accuracy. While Fig.~\ref{fig:MG_contributions} refers to $\Lambda$CDM, we have performed the same analysis for all the MG models listed in  Tab.~\ref{tab:models_fiducial}, and found negligible differences. 

Comparing the number counts in D-space (GWs) with those in z-space (galaxies), we observe similar trends, with some crucial differences involving primarily lensing. We find that the net lensing convergence contribution is always positive in D-space at the scales considered, while in z-space the lensing curve at $z=1.5$ presents the well known V-shaped (divergent) feature, which happens when the fractional contribution flips sign.
The DSD (RSD) contribution, on the contrary, is always positive in both spaces, and at all redshifts has a similar impact on the signal for both GWs and galaxies; the remaining perturbations weights remain small even for the galaxies, rarely raising above $1\%$, with another relevant difference in the Doppler term, whose impact at intermediate and high redshifts is slightly larger in D-space.

\subsection{Impact of the binning choice}
\label{subsec:binning}

Fixing the free parameters in Eq.~\eqref{eq:gaussian_window} corresponds to making a tomographic binning choice, with $z^i$ being the center of the bin, and $\sigma^i_z$ its width. In a forecast analysis these parameters can be chosen with a certain degree of arbitrariness, though with the caution that the lower value of $\sigma^i_z$ (or the corresponding $\sigma^i_D$) is limited by the luminosity distance measurement uncertainty. This will, in general, depend on the specific GW/SN survey considered, though it can impact the results significantly. In general, wider bins have the effect of smoothing out the number count correlation. However, they can also impact differently the individual components of the signal.

\begin{figure}
    \centering
    \includegraphics[width=1.\textwidth]{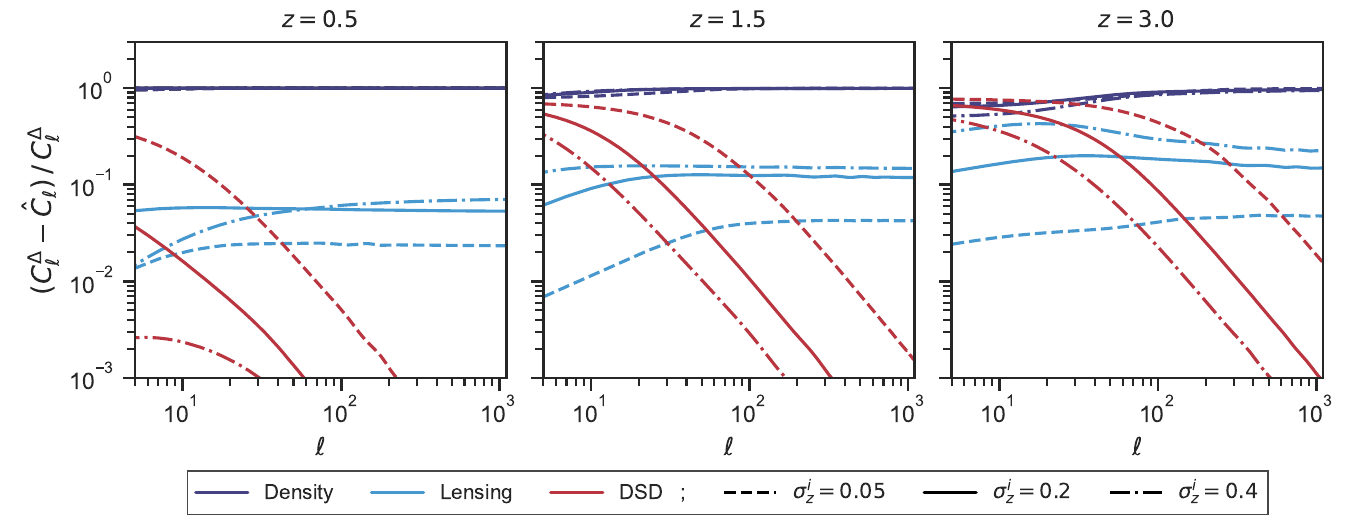}
    \caption{Relative contributions to the total signal (same as Fig.~\ref{fig:MG_contributions}) of the density contrast, lensing and DSD. We plot these contributions at different redshifts and for different choices of the tomographic bin width $\sigma_z^i$. We notice that the binning choice has an important impact in determining the relative weight of each perturbation within the number count, with density contrast and DSD dominating thinner bins, while lensing becomes more and more relevant for wider bins, as both the density contrast and the DSD average out faster.}
    \label{fig:binning}
\end{figure}

We illustrate the effects of the binning choice in Fig.~\ref{fig:binning}, where we plot again the fractional contributions to the total correlation, this time focusing on the dominant terms, density, lensing  and DSD,  and for different choices of the bin width. 
We can appreciate that, for any binning choice, the density contrast is always the dominant ingredient of the clustering signal, with the only exception already pointed out for Fig.~\ref{fig:MG_contributions} of high redshifts and low multipoles. However, in the case of fine binning, the impact of the DSD on the total correlation becomes more and more important, coming to dominate over the lensing contribution at all scales at high $z$, and at $\ell \leq 50,\,300$ for $z=0.5, \, 1.5$.
On the contrary, for wider bins the DSD smooth out extremely fast, letting lensing become the second most important component at all scales above $\ell\geq10$. For this binning choice, lensing becomes particularly relevant also at high redshifts and very low scales, where its relative weight is comparable to that of the DSD and of the density contrast.

As a final remark, we stress that the quantities plotted in Fig.~\ref{fig:binning} are relative weights. Thus, the fact that the lensing curves increase for larger bins must not be deceiving: all correlations, and lensing in particular, are partially averaged out over large bins, i.e. all absolute signals are decreasing. What is changing is their relative importance within the total number count correlation, with lensing becoming more and more important as DSD and density contrast average out {\it faster}.

We conclude that, in general, the binning choice has a huge impact on the individual components, and this should be taken into account carefully in forecast analyses. We note that the reference choice of $\sigma^i_z=0.2$ that we used for most of the paper is compatible with a luminosity distance measurement uncertainty of $\sim 10\%$ or less. This is consistent with several future observational scenarios, but it might be too narrow for some of the more conservative setups (see discussion in Sec.~\ref{sec:observations}).

\subsection{GW number count in ST theories: deviations from \lcdm{}}

\begin{figure}
    \centering
    \includegraphics[width=1.\textwidth]{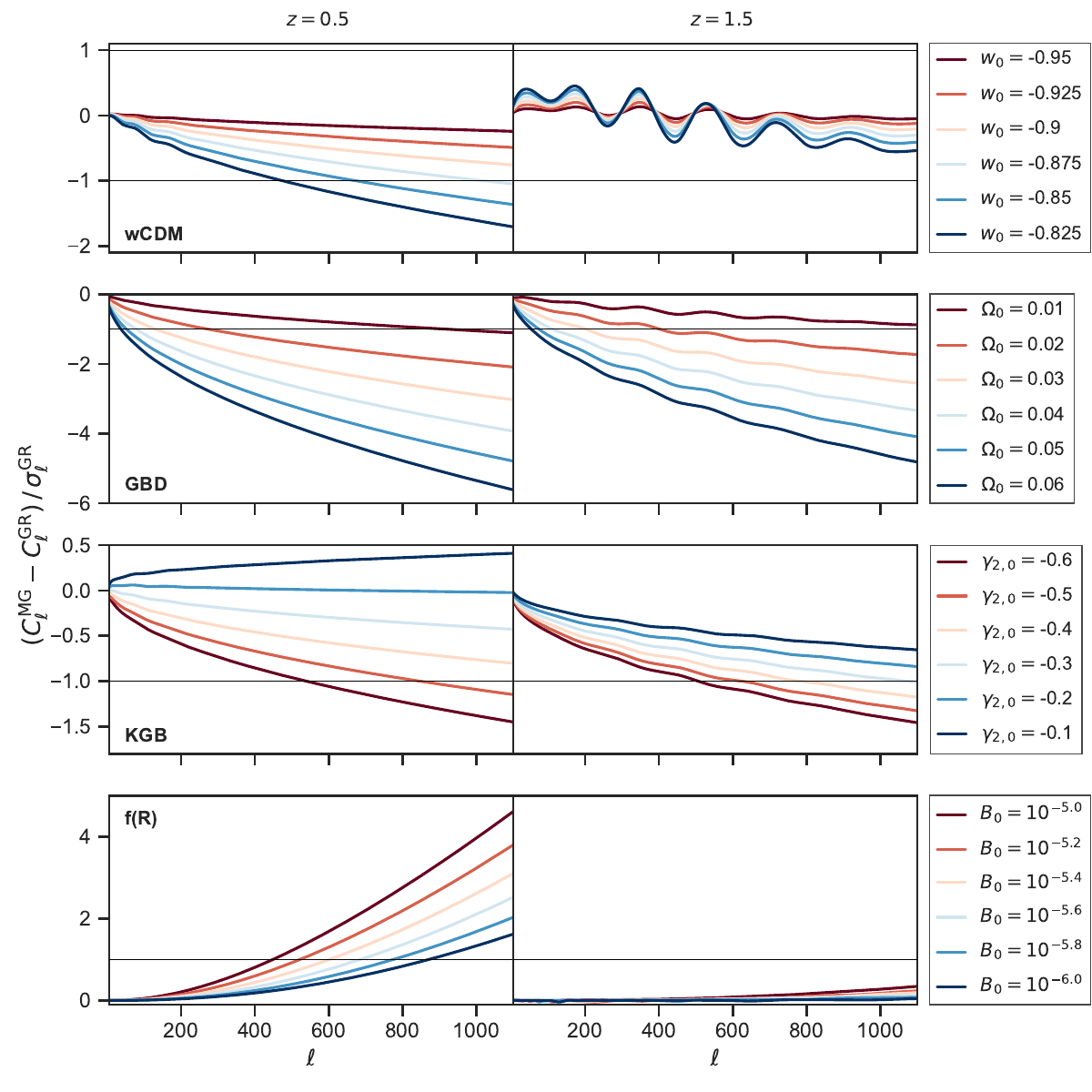}
    \caption{Deviation from \lcdm{} of the auto-correlation of GW clustering for different MG theories, in units of cosmic variance. The theories explored are, from top to bottom, $w$CDM, GBD, KGB and f(R). The different curves are obtained varying the theories parameters. Respectively, we varied: $w_0$ in the interval $[-0.95,-0.825]$ in the first row, $\Omega_0$ in $[0.01, 0.06]$ in the second row, $\gamma_2$ in the interval $[-0.6,-0.1]$ in the third row and $B_0$ in $[10^{-6}, 10^{-5}]$ in the bottom row. The remaining parameters for each theory were left fixed at the fiducial values of Table~\ref{tab:models_fiducial}.}
    \label{fig:MGresults}
\end{figure}

To better highlight the effects of MG on the GW number count,  in Fig.~\ref{fig:MGresults} we plot the deviations from GR of the angular correlations in units of cosmic variance (referred to GR), i.e.
\begin{equation}
    \frac{\Delta C_{\ell}}{\sigma^{\rm GR}_{\ell}} \equiv \sqrt{\frac{2\ell + 1}{2}} \frac{(C_{\ell}^{\rm MG} - C_{\ell}^{\rm GR})}{C_{\ell}^{\rm GR}}\,,
\end{equation}
for the models  described in sub-section~\ref{STmodels} .

 Any deviation higher than $1\, \sigma^{\rm GR}_{\ell}$ could potentially be observed, provided that the right multipoles are accessible for observation, and that the measurement error is small enough. We highlight with black horizontal lines the corresponding limits, i.e. ${\Delta C_{\ell}}/{\sigma^{\rm GR}_{\ell}}=\pm1$,  in the Figure. 
For $w$CDM models,  we keep $w_a$ fixed at $-0.05$, while we vary $w_0$ in the range $[-0.825, -0.95]$. These values are all compatible within $1\sigma$ with the latest DES Year 3 bounds and DES+SN joint constraints, and within $2\sigma$ of the joint DES+BAO+BBN and DES+Planck bounds~\cite{DES:2021wwk}.
We see that a simple modification of the dark energy equation of state does not induce big deviations in the correlation signal with respect to the GR case. Both at low and at high redshifts, these deviations almost always remain within the cosmic variance limit. Only deviations induced by $w_0\leq-0.85$ could be detectable at low redshifts, though the measurement requires a high level of accuracy and the access to multipoles higher than 400. Exploring the GBD case,  we find that this is one class of models that, in general, induces the highest deviations in the correlation curves. We keep $w_0$ and $w_a$ fixed to their fiducial values, while we vary the value of the non minimal coupling today, $\Omega_0$. We observe that for any value of $\Omega_0 > 0.01$, deviations from \lcdm{} are potentially detectable, provided that the highest multipoles are available. In particular, a value of $\Omega_0 \geq 0.02$ generates deviations from GR that can be detectable both at $z=0.5$ and $z=1.5$, for $\ell \geq 300$ and $\ell \geq 400$ respectively, while the deviation signal itself can reach up to $5 \, \sigma^{\rm GR}_{\ell}$. All values of $\Omega_0$ explored here are consistent within $3\sigma$ with the bounds of~\cite{Frusciante:2018jzw} which are, to the best of our knowledge, the most recent constraints put on the specific parametrization of the EFT functions we adopted in this work. In particular, $\Omega_0=0.02$ is closest to their best fit value for the GBD model, $\Omega_0 = 0.018$. 

For KGB models, we keep fixed $w_0, w_a, \Omega_0$ and $\gamma_1^0$. Past works (see e.g.~\cite{Frusciante:2018jzw}) have shown that LSS and CMB depend very mildly on  $\gamma_1(\eta)$ and we have no reason to believe that GW would be more sensitive to this EFT function\footnote{GWs trace the same perturbations, and $\gamma_1$ does not enter in the explicit MG contributions coming through $\alpha_M$ in $\beta$ and $\gamma$ (see Sec.~\ref{sec:number_count_st}).}, (in fact, we checked that that is the case). 
We therefore vary  only on $\gamma_{2,0}$,  in the interval $[-0.6, -0.1]$ which is compatible with the bounds of~\cite{Frusciante:2018jzw}. We  find small deviations from the GR case. The deviations are particularly difficult to capture at low redshifts, with only values of $\gamma_{2,0}\leq-0.5 (0.6)$ give a detectable signal for multipoles higher than 600(800). They become slightly more significant at $z=1.5$, giving a detectable signal for $\gamma_{2,0}\leq -0.4$, provided that $\ell \geq 500 $ are observable.

Finally,  for the $f(R)$ family, we yet again fix the background and vary only  $B_0$. Like GBD, $f(R)$ models give potentially detectable deviations from GR in the GW number counts, some of which already higher than $1\sigma{}_\ell$ for $\ell \geq 400$. However, the relevance of these deviations is limited to low redshifts, while at $z=1.5$ the signal is essentially indistinguishable from \lcdm{} as the MG effects quickly fade. We also notice that $f(R)$ and two of the KGB scenarios are the only cases in which the modifications of gravity actually increase the GW clustering signal. In all other cases explored here, $\Delta C_{\ell}$ is negative, which means that the MG signal is lower than the fiducial GR one. 

Let us note that the deviations shown in  Fig.~\ref{fig:MGresults}  correspond to values of the EFT parameters that are still loosely compatible with the observational bounds placed in~\cite{Frusciante:2018jzw, Raveri:2014cka, Peirone:2016wca} using CMB data and state of the art galaxy surveys. As such,  they are models that in general depart very little from GR. If we were to explore the GW signal separately, as an independent cosmological probe, in principle we could loosen the priors on these parameters and  possibly explore higher departures from GR.

\subsection{Synergy between GW and EM tracers}
\label{sec:synergies}

As already discussed, photons and GWs respond differently to the underlying theory of gravity, with GW geodesics being explicitly affected by the non-minimal coupling and  perturbations in the scalar field. This introduces an interesting difference between the number count in luminosity distance space of SN and that of GWs in the case of ST theories. This comes in addition to the differences due to working in luminosity distance space as opposed to the redshift space, commonly used for galaxies. In this Section we perform a preliminary exploration of the joint constraining power of these three tracers.

\begin{figure}
    \centering
    \includegraphics[width=1.\textwidth]{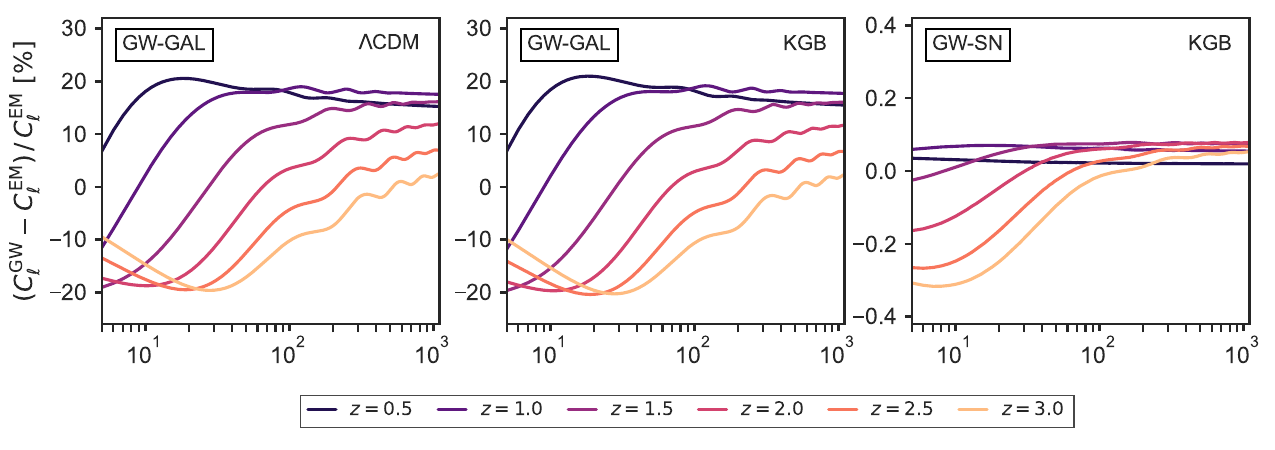}
    \includegraphics[width=1.\textwidth]{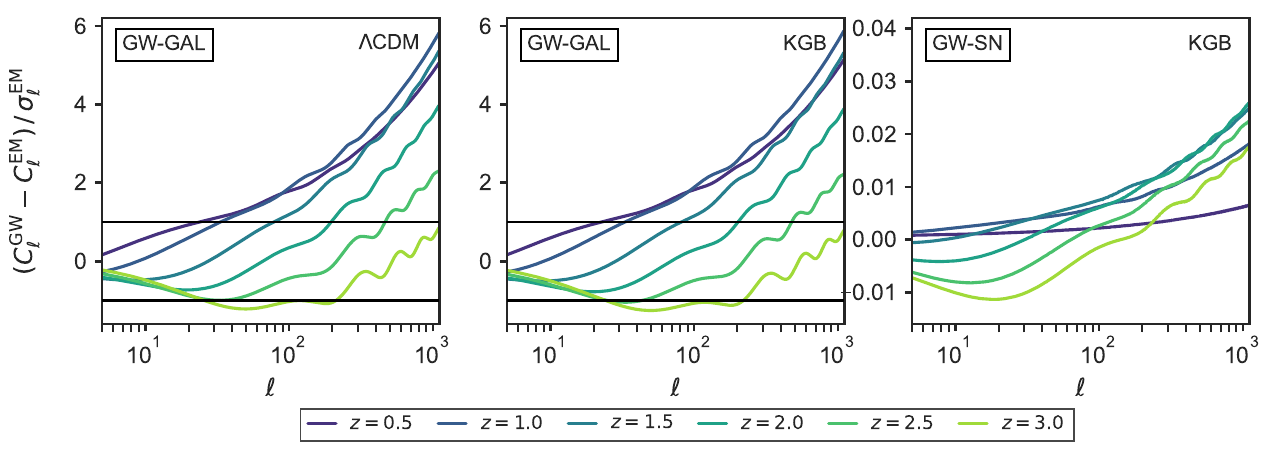}
    \caption{Top row: Relative difference between GW and EM sources number count auto-correlations $(C^{\rm GW}_{\ell} - C^{\rm EM}_{\ell})/C^{\rm EM}_{\ell}$, with ${\rm EM} = [{\rm GAL}, {\rm SN}]$,  at different redshifts in the range $[0.5, 3.0]$. We plot the difference between the GW and the galaxy signal in \lcdm{} (left panel) and in KGB (central panel), and the difference between GW and SN auto-correlations in KGB (right panel).
    Bottom row: same as above, but this time the difference between the auto-correlations $\Delta C_{\ell}$ is plotted in units of cosmic variance $\sigma^{\rm EM}_{\ell}$ and the black horizontal lines mark $\Delta C_{\ell} = \pm \, \sigma^{\rm EM}_{\ell}$. This shows that, even if the percentage deviations between the GW and the EM signal is slightly higher at lower multipoles, it is actually easier to capture it at higher $\ell$, where the cosmic variance noise is smaller.}
    \label{fig:synergy_GW-EM}
\end{figure}

In Fig.~\ref{fig:synergy_GW-EM}, we plot the difference between the GW number count angular correlations and the analogous EM ones, $\Delta C_{\ell}^{\rm GW-EM} = C_{\ell}^{\rm GW} - C_{\ell}^{\rm EM}$, where EM stands for either galaxies (GAL) or supernovae. We focus on $\Lambda$CDM and the KGB model, which among the different ST scenarios that we explore is the one allowing for larger values of $\Omega_0$ (and $\alpha_M$). This is particularly useful for the specific comparison we perform in this Figure, since  a bigger $\alpha_M$ implies a bigger $\Delta C_{\ell}^{\rm GW-SN}$.
We can appreciate the potentialities of a joint galaxies-GW analysis, as $\Delta C_{\ell}^{{\rm GW- GAL}}$ reaches up to $20 \%$ at lower multipoles, plateauing at high $\ell$ around a fixed value between $5$ and $20\%$ for all redshifts below $z=3.0$, and for both models. On the contrary, the difference between GW and SN auto-correlations remains very small, reaching peaks of up to $0.3\%$ at lower multipoles, and settling around $0.05-0.1\%$ for $\ell \geq 100$. Whether these differences are big enough to add valuable cosmological information is better shown in the bottom row of Fig.~\ref{fig:synergy_GW-EM}, where we plot the same quantity $\Delta C_{\ell}^{\rm GW-EM}$, but in units of cosmic variance $\sigma^{\rm EM }_{\ell}$. We see that in the case of SN, $\Delta C_{\ell}$ is only a small fraction of the cosmic variance, indicating that the cross-correlations between GW and SN is unlikely to add any insight about the specific MG model - though we note that a joint analysis would still be worthwhile, as their cross-correlation adds more strength to the clustering signal they both probe independently. On the contrary, the galaxy signal is clearly distinguished from the GW one for all redshifts below $z=3.0$, differing of up to $5-6 \, \sigma^{\rm GAL}_{\ell}$ at high redshifts and high multipoles, hinting that their cross correlation is likely rich of extra cosmological information. In general,  we conclude again that the access to small scales is key, as it is easier to distinguish between GW and EM signals at low z and high multipoles, where the cosmic variance itself is much smaller. We also note, however, that at $z\sim1.0$ it is already possible to distinguish the two signals for $\ell \geq 30$.

The  fact that $\Delta C_{\ell}^{\rm GW-SN}$ is always lower than the $C_{\ell}^{\rm SN}$ cosmic variance means that, as far as the number count cross-correlations are concerned, these two sources give, essentially, the same signal. However, for the sake of completeness, we mention that there is another possibility that concerns these number counts, i.e. the definition of a new observable
\begin{equation}
    \hat{\Delta}_{\rm GW-SN} (D, \hat{n}) \equiv \hat{\Delta}_{\rm GW} (D, \hat{n}) - \hat{\Delta}_{\rm SN} (D, \hat{n}) \, .
\end{equation}
Using Eq.~\eqref{eq:number_count_final} and the analogous one for SN, the full theoretical expression for $\hat{\Delta}_{\rm GW-SN} (D, \hat{n})$ reads
\begin{equation}\begin{split}\label{eq:gw-sn_signal}
    \hat{\Delta} (D, \hat{n}) = & \,\,\, \delta_{\rm GW}^{\rm N} + \int_0^{\bchi} d\chi' \bigg[\frac{\Delta\beta}{2} \frac{\bchi - \chi'}{\bchi\chi'} + \frac{\alpha_M \zeta}{4\H\bchi^2}\bigg] \nabla^2_{\Omega} (\Phi + \Psi) - \frac{\alpha_M \zeta}{\H} \,  \partial_{\bchi} (\bm{v}\cdot\hat{n}) \\[8pt]
    &  - \big( \alpha_M \zeta  + 2 \Delta \beta \big) \bm{v}\cdot\hat{n} + \bigg[\frac{\Delta\beta}{\bchi} + \frac{\alpha_M \zeta}{2\H\bchi^2}\bigg] \int_0^{\bchi} d\chi' (\Phi + \Psi) + 2 \Delta \beta \int_0^{\bchi} d\chi'(\dot{\Phi} + \dot{\Psi}) \\[8pt]
    & + \bigg[\Delta \beta - \frac{\alpha_M \zeta}{2\bchi\H}\bigg] \Phi + \frac{\alpha_M \zeta}{2\H} \big(\partial_{\bchi} \Phi + \dot{\Phi}\big) + \bigg[\frac{\alpha_M\zeta}{2\H\bchi} + 2 \Delta \beta \bigg] \Psi +\\[8pt]
    &+ \gamma \frac{\alpha_M}{2} \frac{d}{d\bar{\eta}}\bigg(\frac{\delta \varphi}{\dot{\varphi}}\bigg) + \bigg[\gamma \frac{\dot{\alpha_M}}{2} + \frac{\alpha_M}{2} \bigg(\gamma \frac{\dot{\H}}{\H}-\beta - 1\bigg) \bigg] \frac{\delta \varphi}{\dot{\varphi}}\,,
\end{split}
\end{equation}
with
\begin{equation}
    \zeta \equiv \gamma^{\rm GW} \gamma^{\rm SN} = \bigg[1 + \frac{1}{\bchi\H} - \frac{\alpha_M}{2}\bigg]^{-1} \bigg[1 + \frac{1}{\bchi\H} \bigg]^{-1}\,,
\end{equation}
and
\begin{equation}\begin{split}
    \Delta \beta \equiv \beta^{\rm GW} - \beta^{\rm SN} =& \frac{\alpha_M}{2} \zeta \bigg[\frac{2}{\bchi{} \H{}} + \frac{\dot{\H{}}}{\H{}^2} - 1\bigg] - \zeta^2 \alpha_M \bigg[1 + \frac{1}{\bchi\H} - \frac{\alpha_M}{4}\bigg] \bigg[\frac{1}{\hchi} \bigg(1 + \frac{\dot{\H}}{\H^2}\bigg)\bigg] \\
    &+\big(\gamma^{\rm GW}\big)^2 \bigg[\frac{\alpha_M}{2} \bigg( 1+ \frac{2}{\hchi}\bigg) - \frac{\alpha_M^2}{4} - \frac{\dot{\alpha}_M}{2\H}\bigg]\,.
\end{split}
\end{equation}
This quantity essentially isolates the explicit MG contribution to the GW number count. In \lcdm{}, $ \hat{\Delta}_{\rm GW-SN} (D, \hat{n})$ is exactly zero, thus the sheer measurement of a non-vanishing value can in principle serve as a smoking gun of physics beyond the standard cosmological model. The practical detection of this quantity might, however, be subject to significant observational challenges. Among these, the request that the number counts of GW and SN need to be observable in the same sky areas down to sufficiently small scales, which might demand unrealistic requirements in terms of numbers, locations and overlap of the two sources (see e.g.~\cite{Garoffolo:2020vtd}). Further explorations in this sense are beyond the scopes of the present work.

\section{Observability of the number count in GW detectors}
\label{sec:observations}

It is interesting to build on the modeling of the previous Sections, to establish what are the observational requirements to carry out number count studies. In this Section, we perform a first step towards this, investigating the possibility of detecting correlations in the number count of compact objects above some noise, for different benchmark observational configurations.

There are several limiting factors affecting the actual detectability of this signal. The experimental error on the luminosity distance determination from the GW amplitude forces a lower limit on the width of the tomographic bins. It thus increases the noise of the measured correlations, while weakening the signal itself, as the perturbative effects partially average out over larger bins (see discussion in Sec.~\ref{subsec:binning}).
On top of this, the total number of detected sources plays a key role, with denser catalogues allowing for a stronger measured correlation. Finally, a severely limiting factor is the uncertainty in the sky-localization of the sources, which reduces the accessible scales cutting out the higher multipoles, where most of the cosmological information is contained. 

To determine whether the number count of Eq.~\eqref{eq:number_count_final} will be observable in future GW detectors, we define a cumulative signal-to-noise ratio (SNR) as
\begin{equation}\label{eq:cumulative_snr}
    \bigg(\frac{S}{N}\bigg)^2_{<\ell} = \sum_{\ell = 2}^{\ell_{\rm max}}  \, \frac{2\ell + 1}{2} \bigg(\frac{C^i_{\ell}}{C^i_{\ell} + N^i_{\ell}}\bigg)^2 \, .
\end{equation}
The noise $N^i_{\ell}$ takes the form of a shot noise dependent on the luminosity distance of the source, and can be written as (see e.g.~\cite{Namikawa:2015prh})
\begin{equation}\label{eq:shot_noise}
    N^i_{\ell} = \frac{4\pi}{N^i_{\rm GW} \sqrt{f_{\rm sky}}} \bigg(\frac{\sigma^i_{\rm GW}}{D^i}\bigg)^2 \, .
\end{equation}
Here, $N^i_{\rm GW}$ is the total number of sources in the $i^{th}$ tomographic bin, while $f_{\rm sky}$ is the sky fraction covered by the survey; for future GW surveys we can set $f_{\rm sky} = 1$, as the proposed  detectors are designed to have essentially no blind spots. $\sigma^i_{\rm GW}$ instead is rigorously defined through
\begin{equation}\label{eq:sigma_bin_def}
    \frac{\sigma^i_{\rm GW}}{D^i} \equiv \int_0^{\infty} dD W^i(D) \, \frac{\sigma_{\rm GW}(D)}{D} \simeq \frac{\sigma_{\rm GW} (D^i)}{D^i} \,,
\end{equation}
where $W^i(D)$ is the bin window function and $\sigma_{\rm GW}(D)$ is the average observational uncertainty on the luminosity distance measurement for a GW source at distance $D$, such that $D^{\rm obs} = D + \sigma_{\rm GW}(D)$. The last equality in Eq.~\eqref{eq:sigma_bin_def} follows from approximating the fractional error $\sigma_{\rm GW}(D)/D$ constant over the bin.

\begin{comment}
\begin{table}
    \renewcommand{\arraystretch}{1.4}
    \centering
    \begin{tabular}{|l|c|c|c|c|c|c|}
        \hline
        Specs. & 3G - Opt. & 3G - Pess. & LISA - Opt. & LISA - Pess. & DO - Opt. & DO - Pess. \\
        \hline
        \hline
        $\sigma_{\rm GW}(D) / D$ & $30\%$ & $10\%$ & $10\%$ & $1\%$ & $1\%$ & $0.5\%$ \\
        $N^{i}_{\rm GW}$ & $10^4$ & $5 \cdot 10^4$ & $10^3$ & $10^3$ & $5 \cdot 10^4$ & $10^5$ \\
        $\ell_{\rm max}$ & $100$ & $100$ & $180$ & $180$ & $1000$ & $1000$ \\
        $\sigma^{\rm bin}_D / D$ & $30\%$ & $10\%$ & $10\%$ & $10\%$ & $10\%$ & $10\%$ \\
        \hline
    \end{tabular}
    \caption{\abc{TBA}}
    \label{tab:detector_specifications}
\end{table}    
\end{comment}

\begin{table}
    \renewcommand{\arraystretch}{1.4}
    \centering
    \begin{tabular}{l|c|c|c|c|c|c}
        \hline
        \hline
        \multirow{2}{*}{Specs.} & \multicolumn{2}{c|}{3G} & \multicolumn{2}{c|}{LISA} & \multicolumn{2}{c}{DO} \\ \cline{2-7}
                                & Cons. & Opt. &
                                Cons. & Opt. & Cons. & Opt. \\
        \hline
        $\sigma_{\rm GW}(D) / D$ & $30\%$ & $5\%$ & $10\%$ & $1\%$ & $1\%$ & $0.5\%$ \\
        $N^{i}_{\rm GW}$ & $10^4$ & $5 \cdot 10^4$ & $10^3$ & $10^3$ & $5 \cdot 10^4$ & $10^5$ \\
        $\ell_{\rm max}$ & $180$ & $180$ & $180$ & $180$ & $1000$ & $1000$ \\
        $\sigma^{\rm bin}_D / D$ & $30\%$ & $10\%$ & $10\%$ & $10\%$ & $10\%$ & $10\%$ \\
        \hline
        \hline
    \end{tabular}
    \caption{Specifications for the benchmark observational scenarios explored in this work. In particular, we set the average fractional uncertainty on the luminosity distance measurement $\sigma_{\rm GW}(D_i)/D_i$ to be used for the shot noise calculation, assumed constant at all redshifts; the number of compact objects detected in each tomographic bin $N^i_{\rm GW}$; the maximum multipole accessible $\ell_{\rm max}$; and the width $\sigma^{\rm bin}_D$ of the luminosity distance tomographic bin. Specifications are chosen to be loosely compatible with expected observational setups  for 3rd generation ground based detectors, a LISA-like space based interferometer and a space based Deci-Hz observatory, in their pessimistic and optimistic configurations.}
    \label{tab:detector_specifications}
\end{table}

The values for the observational uncertainties in Eq.~\eqref{eq:cumulative_snr} need to be chosen relying on forecasts for the next generation detector performances. We consider three different classes of detectors: the 3rd generation (3G) ground based interferometers, a LISA-like space based interferometer and the deci-Hertz space based observatories. The next generation of ground based detectors, such as the Einstein Telescope (ET,~\cite{Maggiore:2019uih}) and Cosmic Explorer (CE,~\cite{Reitze:2019iox}), will be most sensitive to the high frequency band ($\sim 1-10^4 \, {\rm Hz}$), where there is a richness of inspiraling and merging stellar mass binaries (SMB). Based on the current rate of events measured by the LIGO-Virgo-Kagra network~\cite{LIGOScientific:2020kqk, KAGRA:2021duu}, forecasts indicate that  the number of sources that could be detected in 10 years of data taking by ET is as high as $O(10^5)$ for binary black holes (BBH) and neutron star-black hole binaries (NSBH), and $O(10^6)$ for binary neutron stars (BNS)~\cite{Maggiore:2019uih, Pieroni:2022bbh, Ronchini:2022gwk}, while the number and quality of the measurements improve hugely if the observation is carried out jointly by a network of ET and one or two CE~\cite{Ronchini:2022gwk, Belgacem:2019tbw, Chan:2018csa, Zhao:2017cbb, Vitale:2016icu}. 
 We select two scenarios for the 3G detectors: a \emph{conservative} scenario in which we assume a measurement error on the luminosity distance of $30\%$ and require at least $10^4$ sources in each tomographic bin; and an \emph{optimistic} scenario, with $\sigma_{\rm GW}(D)/D = 5\%$ and $N^i_{\rm GW} = 5 \cdot 10^4$. We stress that in both cases, the number of sources that we put in each redshift bin is $1$ or even $2$ orders of magnitude lower than the total number of expected detections, i.e. we are implicitly considering the (realistic) case in which the requested accuracy on the luminosity distance is met only by a fraction of the detected sources.

Next generation space-based detectors such as the Laser Interferometer Space Antenna (LISA,~\cite{LISA:2017arv}), Taiji~\cite{Taiji:2017nwx} and TianQin~\cite{TianQin:2020hid} will be sensitive to low frequency GW ($\sim 10^{-4} - 0.1 \, {\rm Hz}$). Their extra-galactic targets will primarily be massive black holes binaries (MBHB) and extreme mass-ratio inspirals (EMRIs) systems. Though the merger rates for these sources are still highly uncertain, even in the most promising astrophysical scenarios LISA is not expected to measure more than a few hundreds MBHB over the whole 4 years mission duration ~\cite{LISA:2022yao}. On the contrary, EMRIs are potentially much more numerous, with rates ranging between $\sim O(10^{2-3})$ detections per year, depending on the astrophysical model~\cite{LISA:2022yao, Babak:2017tow}. LISA will provide excellent luminosity distance measurements for these sources, thus for LISA-like detectors we consider an optimistic scenario, where we choose $\sigma_{\rm GW}(D)/D = 1\%$ and a more conservative scenario, with $\sigma_{\rm GW}(D)/D = 10\%$. In both cases, we impose $N^i_{\rm GW} = 10^3$. As such, both LISA-like scenarios are to be regarded as upper limits, as they require, to begin with, a high number of detections. We note that if these numbers are not met, it is unlikely that LISA detections can be used efficiently for number counts studies. We further note that the LISA cases can also be representative of two corresponding 3G scenarios, in which a few thousands of golden SBH are detected with extremely accurate luminosity distances. 

Further away in time, the frequecy gap between LISA and 3G detectors will be covered by space based deci-Hertz interferometers. This band contains, among other signals, the inspirals of millions of SBH. Proposed detectors operating in these frequency range are the Advanced Laser Interferometer Antenna (ALIA,~\cite{Bender:2013cqg, Baker:2019ync}), the Big Bang Observer (BBO,~\cite{Crowder:2005nr}) and the DECi-hertz Interferometer Gravitational wave Observatory (DECIGO~\cite{Sato:2017dkf, Kawamura:2020pcg}). While the actual technical feasibility of these missions is under investigation, a slightly less ambitious family of detectors in the deci-Hertz band, generically named Deci-hertz Observatories (DO) has been proposed in ~\cite{Sedda:2019uro}. Given the early planning stages of these experiments, solid forecasts on the measurement accuracy are unavailable, but based on preliminary estimates (see e.g.~\cite{Crowder:2005nr}), they are likely to outperform both LISA and 3G observatories. Tentatively, for this family of detectors we again consider a conservative and an optimistic scenario, requiring $5 \cdot 10^4$ and $10^5$ detections in each bin, and $\sigma_{\rm GW}(D)/D = 1\%$ and $0.5\%$ respectively.

We summarize the assumed specifications for each family of detectors in Tab.~\ref{tab:detector_specifications}. While we vary the accuracy on the luminosity distance measurement to be used in Eq.~\eqref{eq:shot_noise} for the shot noise calculation, we keep fixed the relative gaussian bin width to $\sigma^{\rm bin}_D/D^i = 10\%$, with the only exception of the 3G conservative scenario, where the higher uncertainty on the distance measurements imposes wider bins. This choice is made consistently with the requirement of a high number of sources within each bin, while assuming constant ratios for $\sigma^i_{\rm GW}/D^i$ and $\sigma^{\rm bin}_D/D^i$ automatically accounts for a redshift evolution in the luminosity distance uncertainty, and consequently of the bin width.

\begin{figure}
    \centering
    \includegraphics[width=1.\textwidth]{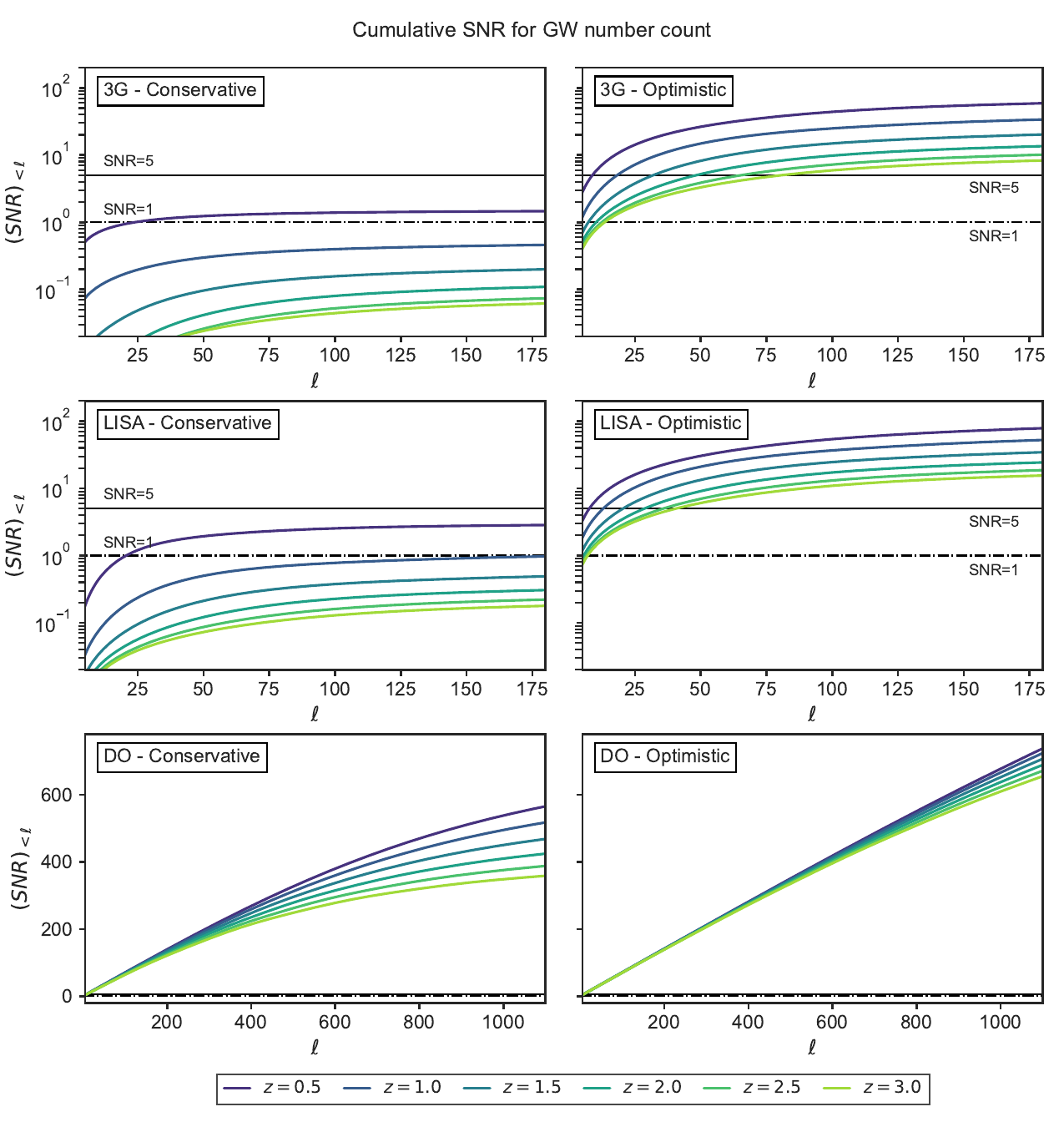}
    \caption{Cumulative SNR for the GW number count correlation, at different redshifts. The observational scenarios considered in terms of accuracy of the luminosity distance measure and number of detected sources are those of Tab.~\ref{tab:detector_specifications}, specifically: a network of ground based 3G detectors (first row), a LISA like space based interferometer (second row) and a DO space based observatory (bottom row). Gaussian bins widths are fixed at $\sigma_{\rm bin}/D^i = 10\%$ ($30\%$ for the 3G - Conservative scenario). The dash-dotted and solid horizontal lines in each panel indicate a cumulative SNR of 1 and 5 respectively.}
    \label{fig:SNR}
\end{figure}

In Fig.~\ref{fig:SNR} we plot the cumulative SNR for the different scenarios described above, and for different redshifts, clearly marking the cumulative SNR of $1$ and $5$. We plot all 3G and LISA curves up to $\ell_{\rm max} = 180$. This value corresponds, roughly, to an average sky-localization of $1 \, {\rm deg}^2$ for each source. Better localizations are extremely unlikely to be achieved with these detectors, especially for a high number of sources, and indeed though the $1 \, {\rm deg}^2$ limit might be met with BBH observations~\cite{Zhao:2017cbb}, BNS detection are expected to be way more uncertain, with only $\sim 20 \%$ of detected source localizaed within $10 \, {\rm deg}^2$~\cite{Ronchini:2022gwk}, roughly corresponding to $\ell_{\rm max} = 50$. The DO family instead is way more likely to achieve optimal sky-localization. Expectations depend massively on the number of interferometers that can be put in a network in space, as more detectors allow a better triangulation of the GW. However, it is expected that a DO-like detector alone could measure signals up to $\ell_{\rm max} \sim 300$~\cite{Sedda:2019uro}, while a network of DO could localize each source within a few ${\rm arcmin}^2$~\cite{Crowder:2005nr}, i.e. $\ell_{\rm max} \geq 1000$.

From Fig.~\ref{fig:SNR}, we can see that for both LISA and 3G configurations, only the optimistic case produces a number count correlation signal with SNR $\geq 5$. For the conservative scenarios, a correlation is detectable only at low redshifts ($z\leq 0.5$) and with very low significance (SNR $\leq 1.5$). The optimistic configurations instead produce angular correlations that are detected almost at all multipoles, even with a $5\sigma$ significance for the smallest redshifts. However, for redshifts $z=2$ and higher,  detecting a correlation signal  requires access to the smallest scales, $\ell \geq 70$ for 3G detectors and $\ell \geq 60$ for LISA. Going up to $\ell = 100$, however, could produce a cumulative SNR of $~50-70$ for the smallest redshifts. This is once again proof that optimal source localization will be key to determine the feasibility of number count studies with the nest generation of detectors. For the DO case instead, both explored scenarios give optimal SNR: the signal is detectable with $5 \sigma$ significance at all multipoles, and even at multipoles around $\ell \sim 100$, the cumulative SNR is higher than $100$ at essentially all redshifts. A network of space based DO could probe scales up to $\ell \sim 1000$, accumulating an SNR of several hundreds.

As a final remark, we stress that there are additional sources of errors that we did not account for here, among which a redshift dependent GW source distribution and the completeness of the detected GW catalogues. These  can  impact especially the high redshifts, where sources are fewer and more difficult to detect. We partially compensate for the former using wider bins at higher redshifts, although a more accurate determination of its influence might further lower the expected SNR at $z \geq 2.0$. However, for lower redshifts we do not expect these effects to have a significant impact on our conclusions. 

\section{Summary and conclusions}
\label{sec:conclusions}

We have derived the theoretical expression for the number count of objects living in luminosity distance space,  in the broad context of scalar-tensor theories of gravity with luminal propagation of the tensors, including the full set of relativistic effects. We have shown that, contrary to the GR case, the theoretical form of the number count differs if the objects considered are gravitational waves (GW) or Supernovae (SN). This, because the propagation of GW waves is perturbed differently than that of photons in the presence of a non-minimal coupling with the scalar field.

We have investigated the impact of relativistic effects on the number count in luminosity distance, for both GW and SN, by mean of the 2-point angular correlation function associated to it. Similarly to the galaxy case, we find that the effect that dominate the signal at essentially all redshifts and scales is the density contrast - which is probed directly from the number count. Following suit, the two most important effects on top of that are the weak lensing convergence and the luminosity distance space distortions (DSD). For the latter, we have explored how  their contribution to the total signal depends on the  choice for the tomographic bins.  
We have also shown that all other perturbations remain sub-dominant with respect to lensing and DSD, with Doppler contributing at most $O(1\%)$ of the correlation signal, similarly to all potential terms, time delay, ISW and the scalar field clustering combined.

Focusing on scalar-tensor models, we have explored the deviations of the GW number count angular correlations from the GR counterpart. With the spirit of investigating the potential for detectability, we have plotted these deviations in units of cosmic variance $\sigma^{\rm GR}_{\ell}$. While departures from \lcdm{} are small for all models, the theories that give the most promising signatures are GBD and $f(R)$, for which they can reach up to $6 \, \sigma^{\rm GR}_{\ell}$, especially at low redshifts. On the contrary, for the class of KGB and $w$CDM models departures from GR are tiny, and barely above $1 \, \sigma^{\rm GR}_{\ell}$ only for the most extreme values still allowed for the MG parameters.

We have also investigated the  synergy of GW with other EM tracers, either living in redshift space (galaxies) or in luminosity distance space (SN). 
In \lcdm{} and in alternative theories, we find that the number count of GW and galaxies give two very well distinguished signals for $z<3.0$, differing by up to $25 \%$ and up to $5-6 \, \sigma^{\rm GAL}_{\ell}$; therefore, their cross-correlation can add valuable information in cosmological analyses. On the other hand, despite the theoretical differences in the GW and SN number counts, the respective auto-correlations differ, at most, by  fractions of the SN cosmic variance $\sigma^{\rm SN}_{\ell}$.
However, we show that it is still possible to additionally {\it define} a new observable, in terms of the difference between the SN and GW number count. This quantity isolates the MG contributions to the number count, while being zero in \lcdm{}, thus its detection could serve in principle as a smoking gun of exotic physics. We give the theoretical expression for this observable, though we strongly stress that the actual feasibility of its detection is yet to be proven.

We have also explored the potential observability of the GW number count in planned ground-based and space-based observatories, concluding that it can be detected with the next generation of interferometers only in the optimal scenarios, especially at high redshifts. Crucial for the observation are a high statistic of detected mergers, a low uncertainty on the luminosity distance determination and the access to the lowest scales, i.e. a good angular localization of the sources. We find the former factor to be the biggest limitation for a LISA-like detector, while the last two affect more heavily the ground-based ET and CE. However, for the DO family, we find that the GW number count is detectable at all redshifts, with SNR that can reach several hundreds, likely allowing for optimal cosmological analyses. 
 We conclude that, in the long run, GW have the potentialities to become an extremely effective probe to test the full extent of the cosmological model.

As a final remark, we highlight that there are further observational uncertainties that we have not considered in this work. Likely the most impactful of them is the GW magnification bias, for which we only provided the theoretical expression in ST theories. We have shown that for GW this bias is model-dependent, as the number of sources that are detected above a certain SNR threshold depends on the background cosmology {\it and} on the value of the non-minimal coupling. We deferred the full modeling of these effects to a future investigation. However, we caution that, given the importance that the magnification bias has for galaxy surveys (see e.g.~\cite{Duncan:2021jxl, Thiele:2019fcu}), we expect it to be as important for GW, and to be an essential ingredient in forecast analyses.

\acknowledgments
We thank Alice Garoffolo, Daniele Bertacca, Nicola Bartolo, Tessa Baker, José Fonseca and Stefano Zazzera for fruitful discussions on the luminosity distance perturbations. We are also grateful to Marco Raveri and Fabrizio Renzi for useful suggestions on the {\tt EFTCAMB} implementation. AB acknowledges support from the de Sitter Fellowship of the Netherlands Organization for Scientific Research (NWO). AS and MP acknowledge  support from the NWO and the Dutch Ministry of Education, Culture and Science (OCW) (through NWO VIDI Grant No. 2019/ENW/00678104 and ENW-XL Grant OCENW.XL21.XL21.025 DUSC) and from the D-ITP consortium.
\appendix

\section{Full derivation of the perturbations}
\label{app:perturbations}

In this Appendix, we show the full calculation that leads from Eq.~\eqref{eq:counts_expanded} to Eq.~\eqref{eq:counts_general}. In doing so, we follow very closely the calculation of~\cite{Bonvin2011}, though we specify immediately to Newtonian gauge. We still report our calculation fully for completeness, as some of the resulting equations are presented in a more general form than done in~\cite{Bonvin2011}, with the intent of generalizing the result to cosmological models beyond \lcdm{}.

\subsection{Density perturbations}
\label{app:density_perturbations}

We begin with the density perturbations
\begin{equation}\label{eq:density_perturbations_def}
    \frac{\rho(D, \hat{n}) - \bar{\rho}(D)}{\bar{\rho}(D)} = \delta^N_{\rm GW} - \frac{1}{\bar{\rho}}\frac{\partial\bar{\rho}}{\partial\bar{D}}\bigg|_{D=\bar{D}} \Delta{D}\,.
\end{equation}

The first term on the RHS is simply the GW clustering in Newtonian gauge. For the second term we use that
\begin{equation}
\bar{\rho}(\bar{D}) \equiv a^{-3} \bar{n}(\bar{D})\,,    
\end{equation}
where $\bar{n}(\bar{D})$ is the physical comoving number density of sources, and we obtain
\begin{equation}\label{eq:density_pert_1}
    \frac{1}{\bar{\rho}}\frac{\partial\bar{\rho}}{\partial\bar{D}} = \frac{1}{\bar{\rho}}\frac{\partial\bar{\rho}}{\partial a} \frac{da}{d\bar{z}} \bigg(\frac{d\bar{D}}{d\bar{z}}\bigg)^{-1} = \frac{1}{\bar{n} a^{-3}} \frac{\partial(\bar{n}a^{-3})}{\partial a} (-a^2) \frac{\gamma}{a \bar{D}} = \bigg[3 - \frac{\partial \ln \bar{n}}{\partial \ln a} \bigg] \frac{\gamma}{\bar{D}}\,,
\end{equation}
having defined $\gamma \equiv a\bar{D} \big(d\bar{D}/d\bar{z}\big)^{-1}$. Plugging Eq.~\eqref{eq:density_pert_1} into~\eqref{eq:density_perturbations_def} gives
\begin{equation} \label{eq:density_perturbations}
    \frac{\rho(D, \hat{n}) - \bar{\rho}(D)}{\bar{\rho}(D)} = \delta^N_{\rm GW} - \bigg[3 - \frac{\partial \ln \bar{n}}{\partial \ln a} \bigg] \gamma \, \frac{\Delta D}{\bar{D}}\,.
\end{equation}

The term $\partial \ln\bar{n}/\partial \ln a$ is the evolution bias of GW, totally analogous to the same term affecting the clustering of galaxies.

\subsection{Volume densities}
\label{app:volume_densities}

Before proceeding to the calculation of the volume perturbations, a very useful quantity that will come in handy is the volume density. This can be defined starting from the infinitesimal volume element
\begin{equation}\begin{split}\label{eq:volume_element}
    dV &= \sqrt{-g} \epsilon_{\mu \nu \alpha \beta} u^{\mu} dx^{\nu} dx^{\alpha} dx^{\beta} = \sqrt{-g} \epsilon_{\mu \nu \alpha \beta} u^{\mu} \frac{\partial x^{\nu}}{\partial D} \frac{\partial x^{\alpha}}{\partial \theta_s} \frac{\partial x^{\beta}}{\partial \phi_s} \bigg| \frac{\partial (\theta_s, \phi_s)}{\partial (\theta_o, \phi_o)}\bigg| \frac{1}{\sin{\theta_o}} dD d\Omega_o =\\
    &\equiv v(D, \theta_o, \phi_o) dD d\Omega_o\,,
\end{split}
\end{equation}
with $d\Omega_o \equiv \sin(\theta_o) d\theta_o d\phi_o$. In~\eqref{eq:volume_element}, $(\theta_s, \phi_s)$ identify the true sky location of the considered volume, while $(\theta_o, \phi_o)$ mark its observed location. $v(D, \theta_o, \phi_o)$ is the volume density defined as
\begin{equation}\label{eq:volume_density_def}
    v(D, \theta_o, \phi_o) \equiv \sqrt{-g} \epsilon_{\mu \nu \alpha \beta} u^{\mu} \frac{\partial x^{\nu}}{\partial D} \frac{\partial x^{\alpha}}{\partial \theta_s} \frac{\partial x^{\beta}}{\partial \phi_s} \bigg| \frac{\partial (\theta_s, \phi_s)}{\partial (\theta_o, \phi_o)}\bigg| \frac{1}{\sin{\theta_o}}\,.
\end{equation}

Then, recalling the definition of our background and perturbed frames in Sec.~\ref{sec:reference_frames}, $(\theta_s,\phi_s)$ always correspond to the physical location of the source $(\btheta, \bphi)$, while $(\theta_o,\phi_o)$ depend on the frame in which the volume is considered. \\

We can compute the background volume density $\bar{v}(\bar{D}, \btheta, \bphi)$ using Eq.~\eqref{eq:volume_density_def}, and noting that, at the background level, $\sqrt{-g} = a^4$, $\theta_o = \btheta$, $\phi_o = \bphi$, $u = 1/a (1,\vec{0})$ and $\bar{x}^i = \bchi{} \bar{n}^i$.\\
Furthermore, we can write
\begin{equation}
    \frac{\partial \bar{x}^i}{\partial \bar{D}} = \frac{\partial \bar{x}^i}{\partial \bchi{}} \frac{d \bchi{}}{d\bar{z}} \bigg(\dfrac{d\bar{D}}{d\bar{z}}\bigg)^{-1} = \frac{\gamma}{\bar{D}\H{}} \bn^i\,,
\end{equation}
while the partial derivatives of the comoving coordinates w.r.t. the angles are
\begin{equation}
    \frac{\partial \bar{x}^i}{\partial \btheta} = \bchi \, \frac{\partial \bn^i}{\partial \theta_s} \qquad \text{and} \qquad \frac{\partial \bar{x}^i}{\partial \bphi} = \bchi \, \frac{\partial \bn^i}{\partial \bphi}\,,
\end{equation}
with
\begin{equation}    
\begin{displaystyle}
  \begin{aligned}
    \frac{\partial \bar{n}^1}{\partial \btheta} &= \cos\bphi \cos\btheta\,, \\
    \frac{\partial \bar{n}^2}{\partial \btheta} &= \sin\bphi \cos\btheta\,, \\
    \frac{\partial \bar{n}^3}{\partial \btheta} &= - \sin\btheta \,,\\
  \end{aligned} \qquad \text{and} \qquad 
  \begin{aligned}
    \frac{\partial \bar{n}^1}{\partial \bphi} &= - \sin\bphi \sin\btheta\,, \\
    \frac{\partial \bar{n}^2}{\partial \bphi} &= \cos\bphi \sin\btheta\,, \\
    \frac{\partial \bar{n}^3}{\partial \bphi} &= 0 \,.\\
  \end{aligned}
\end{displaystyle}
\end{equation}

Because $u^i = 0$ and $\partial \bar{x}^3/\partial \bphi = 0$, in Eq.~\eqref{eq:volume_density_def} only four terms survive, namely
\begin{equation}\begin{split}
    \bar{v}(\bar{D}, \btheta, \bphi) &= \frac{a^3}{\sin\btheta} \bigg[ \frac{\partial \bar{x}^2}{\partial \bar{D}} \frac{\partial \bar{x}^3}{\partial \btheta} \frac{\partial \bar{x}^1}{\partial \bphi} + \frac{\partial \bar{x}^3}{\partial \bar{D}} \frac{\partial \bar{x}^1}{\partial \btheta} \frac{\partial \bar{x}^2}{\partial \bphi} - \frac{\partial \bar{x}^3}{\partial \bar{D}} \frac{\partial \bar{x}^2}{\partial \btheta} \frac{\partial \bar{x}^1}{\partial \bphi} - \frac{\partial \bar{x}^1}{\partial \bar{D}} \frac{\partial \bar{x}^3}{\partial \btheta} \frac{\partial \bar{x}^2}{\partial \bphi} \bigg] =\\
    &= \frac{a^3}{\sin\btheta} \frac{\gamma \bchi{}^2}{\bD\H{}} \bigg[\bn^2 \frac{\partial \bn^3}{\partial \btheta} \frac{\partial \bn^1}{\partial \bphi} + \bn^3 \frac{\partial \bn^1}{\partial \btheta} \frac{\partial \bn^2}{\partial \bphi} - \bn^3 \frac{\partial \bn^2}{\partial \btheta} \frac{\partial \bn^1}{\partial \bphi} - \bn^1 \frac{\partial \bn^3}{\partial \btheta} \frac{\partial \bn^2}{\partial \bphi}\bigg] =\\
    &= \frac{a^3}{\sin\btheta} \frac{\gamma \bchi{}^2}{\bD\H{}} \big[\sin^3\btheta \sin^2\bphi + \cos^2\btheta \sin\btheta \cos^2\bphi  + \sin\btheta \cos^2\btheta \sin^2\bphi + \sin^3\btheta \cos^2\bphi\big] = \\
    &= \frac{a^3}{\sin\btheta} \frac{\gamma \bchi{}^2}{\bD\H{}} \big[\sin\btheta \sin^2\bphi + \sin\btheta \cos^2\bphi\big]\,,
\end{split}    
\end{equation}
which eventually gives
\begin{equation}
    \bar{v}(\bar{D}) = \frac{a^3 \gamma \bchi{}^2}{\bar{D}\H{}}\,.
\end{equation}

As expected, the physical volume density does not depend on the source sky-location, only on its distance $\bD$ from the observer.\\

The calculation of the volume density in the perturbed space is similar to the one in physical space, but with some crucial differences. In the perturbed space we have that $\sqrt{-g} = a^4 (1 + \Psi - 3\Phi)$, $u = 1/a (1-\Psi, \bm{v})$, $\theta_o = \theta$, $\phi_o=\phi$ and the angles are connected, to linear order, by the relation
\begin{equation}\begin{split}\label{eq:angle_transformation}
\btheta &= \theta + \delta \theta\,, \\
\bphi &= \phi + \delta \phi \,,
\end{split}
\end{equation}
which means that the determinant of the Jacobian of the angles transformation now reads
\begin{equation}
    \bigg| \frac{\partial (\theta_s, \phi_s)}{\partial (\theta_o, \phi_o)}\bigg| = 
    \begin{vmatrix}
        1 + \frac{\partial \delta \theta}{\partial \theta} & 0 \\
        0 & 1 + \frac{\partial \delta \phi}{\partial \phi}
    \end{vmatrix}
    = 1 + \frac{\partial \delta \theta}{\partial \theta} + \frac{\partial \delta \phi}{\partial \phi}\,.
\end{equation}

From eq.~\eqref{eq:volume_density_def}, we can split the perturbed volume density in the sum of two terms
\begin{equation}
    v(D,\hat{n}) = v(D,\hat{n})|_{\mu=0} + v(D,\hat{n})|_{\mu = i}\,,
\end{equation}
the calculation of the first term is in all analogous to that of $\bar{v}(\bar{D})$, except now we have a number of prefactors coming from $u^0$, the Jacobian of the angular matrix and $\sqrt{-g}$. We also need to be cautions that all quantities that enter the calculation are perturbed (unbarred) quantities, except the source angles, that are connected to the observed angles through~\eqref{eq:angle_transformation}. Retracing the same steps, we get
\begin{equation}\begin{split} \label{eq:volume_density_1}
    v(D,\hat{n})|_{\mu=0} = &a^4 (1 + \Psi - 3\Phi)\bigg( \frac{1-\Psi}{a}\bigg) \bigg(1 + \frac{\partial \delta \theta}{\partial \theta} + \frac{\partial \delta \phi}{\partial \phi}\bigg) \times \\
    &\times \frac{\chi^2}{\sin\theta} \frac{d\chi}{dD} \bigg[n^2 \frac{\partial n^3}{\partial \btheta} \frac{\partial n^1}{\partial \bphi} + n^3 \frac{\partial n^1}{\partial \btheta} \frac{\partial n^2}{\partial \bphi} - n^3 \frac{\partial n^2}{\partial \btheta} \frac{\partial n^1}{\partial \bphi} - n^1 \frac{\partial n^3}{\partial \btheta} \frac{\partial n^2}{\partial \bphi}\bigg] =\\
    & = a^3 (1 + \Psi - 3\Phi)\bigg[(1-\Psi) \chi^2 \frac{d\chi}{dD} \frac{\sin\btheta}{\sin\theta} \bigg(1 + \frac{\partial \delta \theta}{\partial \theta} + \frac{\partial \delta \phi}{\partial \phi}\bigg) \bigg]\,,
\end{split}
\end{equation}
which gives
\begin{equation}
    v(D,\hat{n})|_{\mu=0} = a^3 (1 + \Psi - 3\Phi)\bigg[\chi^2 \frac{d\chi}{dD} \frac{\sin\btheta}{\sin\theta} \bigg(1 + \frac{\partial \delta \theta}{\partial \theta} + \frac{\partial \delta \phi}{\partial \phi}\bigg) - \Psi \bchi{}^2 \frac{d\bchi{}}{d\bar{D}}\bigg]\,,
\end{equation}
where, in the second term, we have used that $\Psi$ is already a first order perturbation, and we neglected higher order terms.

As for the second term, $v(D,\hat{n})|_{\mu = i}$, considering that $\partial \eta / \partial \btheta = \partial \eta / \partial \bphi = 0$ and that, once again, $\partial x^3 / \partial \bphi = 0$, there are still only 4 terms that survive the contraction of $\epsilon_{i\nu\alpha\beta}$ with the coordinates derivatives. These terms are
\begin{equation}\begin{split}
    v(D,\hat{n})|_{\mu = i} = &a^3 (1 + \Psi - 3\Phi) \bigg(1 + \frac{\partial \delta \theta}{\partial \theta} + \frac{\partial \delta \phi}{\partial \phi}\bigg) \frac{\chi^2}{\sin\theta} \frac{d\eta}{dD} \times \\
    &\times  \bigg[- v^3 \frac{\partial n^1}{\partial \btheta} \frac{\partial n^2}{\partial \bphi} + v^1 \frac{\partial n^3}{\partial \btheta} \frac{\partial n^2}{\partial \bphi} - v^2 \frac{\partial n^3}{\partial \btheta} \frac{\partial n^1}{\partial \bphi} + v^3 \frac{\partial n^2}{\partial \btheta} \frac{\partial n^1}{\partial \bphi}\bigg]\,.
\end{split}
\end{equation}

Now, $\bm{v}$ is already a first order perturbation and as such, since it enters all terms, we can keep only the zeroth order terms in the prefactors and drop the rest. We obtain
\begin{equation}\begin{split} \label{eq:volume_density_2}
    v(D,\hat{n})|_{\mu = i} = &a^3 (1 + \Psi - 3\Phi) \frac{\bchi{}^2}{\sin\theta} \frac{d\bar{\eta}}{d\bar{D}} \bigg[- v^3 \cos^2\bphi \cos\btheta \sin\btheta + \\
    &  - v^1 \sin^2\btheta \cos\bphi - v^2 \sin^2\btheta \sin\bphi - v^3 \sin^2\bphi \sin\btheta \cos\btheta \bigg] = \\
    & = - a^3 (1 + \Psi - 3\Phi) \bchi{}^2 \frac{d\bar{\eta}}{d\bar{D}} \frac{\sin\btheta}{\sin\theta} (\bm{v}\cdot\hat{n}) = - a^3 (1 + \Psi - 3\Phi) \bchi{}^2 \frac{d\bar{\eta}}{d\bar{D}} (\bm{v}\cdot\hat{n})\,.
\end{split}
\end{equation}

Joining Eq.~\eqref{eq:volume_density_1} and~\eqref{eq:volume_density_2}, we are left with the final expression for the perturbed volume density
\begin{equation}\label{eq:perturbed_volume_density}
    v(D,\hat{n}) = a^3 (1 + \Psi - 3\Phi)\bigg[\chi^2 \frac{d\chi}{dD} \frac{\sin\btheta}{\sin\theta} \bigg(1 + \frac{\partial \delta \theta}{\partial \theta} + \frac{\partial \delta \phi}{\partial \phi}\bigg) - \bchi{}^2 \bigg(\Psi \frac{d\bchi{}}{d\bar{D}} + \frac{d\bar{\eta}}{d\bar{D}} (\bm{v}\cdot\hat{n}) \bigg)\bigg]\,.
\end{equation}

\subsection{Volume perturbations}
\label{app:volume_perturbations}

The definition~\eqref{eq:volume_element} allows to rewrite the volume perturbations in terms of perturbations to the volume density. We can write
\begin{equation} \label{eq:volume_perturbations_def}
    \frac{V(D, \hat{n}) - \bar{V}(D)}{\bar{V}(D)} = \frac{v(D, \hat{n}) - \bar{v}(D)}{\bar{v}(D)} = \frac{v(D, \hat{n}) - \bar{v}(\bar{D})}{\bar{v}(\bar{D})} - \frac{1}{\bar{v}} \frac{\partial \bar{v}}{\partial D}\bigg|_{D=\bar{D}}\Delta D\,.
\end{equation}

To compute the first term, we now use the results of the previous sub-section for $v(D,\hat{n})$ and expand at first order all objects in Eq.~\eqref{eq:perturbed_volume_density}. We use that
\begin{equation} \label{eq:sin_ratio}
    \frac{\sin \btheta}{\sin\theta} = \frac{1}{\sin\theta} \sin(\theta + \delta \theta) \simeq \frac{1}{\sin\theta} \bigg[\sin\theta + \cos\theta \delta\theta \bigg] = 1+\cot\theta \delta\theta\,,
\end{equation}
and that $\chi^2 = (\bchi{} + \delta\chi)^2 \simeq \bchi{}^2 + 2\bchi{}\delta\chi$. We also have that
\begin{equation}
    \frac{d\bchi{}}{d\bar{D}} = \frac{d\bchi{}}{d\bar{z}} \bigg(\frac{d\bar{D}}{d\bar{z}}\bigg)^{-1} = \frac{\gamma}{\bar{D}\H{}} \qquad \text{and} \qquad \frac{d\bar{\eta}}{d\bar{D}} = -\frac{d\bchi{}}{d\bar{D}} = - \frac{\gamma}{\bar{D}\H{}}\,.
\end{equation}

Finally, we expand
\begin{equation}
    \frac{d\chi}{dD} = \frac{d\chi}{d\bar{D}} \bigg(\frac{dD}{d\bar{D}}\bigg)^{-1} = \bigg(\frac{d\bchi{}}{d\bar{D}} + \frac{d\delta\chi}{d\bar{D}} \bigg) \bigg(1 - \frac{d\Delta D}{d\bar{D}}\bigg) = \frac{d\bchi{}}{d\bar{D}} + \frac{d\delta\chi}{d\bar{D}} - \frac{d\bchi{}}{d\bar{D}}\frac{d\Delta D}{d\bar{D}}\,,
\end{equation}
that we re-write as
\begin{equation} \label{eq:chi_derivative}
    \frac{d\chi}{dD} = \bigg(-1 - \frac{d\delta\chi}{d\bchi} + \frac{d\bchi{}}{d\bar{D}}\frac{d\Delta D}{d\bchi}\bigg) \frac{d\bar{\eta}}{d\bar{D}}\,,
\end{equation}
where we take the derivatives of the comoving distance perturbations $\delta\chi$ and of the luminosity distance perturbations $\Delta D$ along the null geodetics. Plugging Eqs.~\eqref{eq:sin_ratio}-\eqref{eq:chi_derivative} into Eq.~\eqref{eq:perturbed_volume_density}, we finally obtain
\begin{equation}\begin{split} \label{eq:perturbed_volume_density_final}
    v(D,\hat{n}) &= a^3 (1 + \Psi - 3\Phi) \frac{d\bar{\eta}}{d\bar{D}} \bchi{}^2 \times \\
    &\bigg[ \bigg(1 + 2\frac{\delta\chi}{\bchi{}} \bigg) \bigg(-1 - \frac{d\delta\chi}{d\bchi} + \frac{d\bchi{}}{d\bar{D}}\frac{d\Delta D}{d\bchi}\bigg) \big(1+\cot\theta_o \delta\theta\big) \bigg(1 + \frac{\partial \delta \theta}{\partial \theta} + \frac{\partial \delta \phi}{\partial \phi}\bigg) - \big(\bm{v}\cdot\hat{n} - \Psi \big) \bigg]= \\
    &= \bar{v}(\bar{D}) \bigg[1 + 2\frac{\delta\chi}{\bchi{}} + \bigg(\cot\theta + \frac{\partial}{\partial \theta}\bigg)\delta\theta + \frac{\partial \delta\phi}{\partial \phi} + \bm{v}\cdot\hat{n} - 3\Phi + \frac{d\delta\chi}{d\bchi} -\frac{\gamma}{\bar{D}\H{}}\frac{d\Delta D}{d\bchi} \bigg]\,.
\end{split}
\end{equation}

Then, we need to compute the second term at the RHS of Eq.~\eqref{eq:volume_perturbations_def}, that yields
\begin{equation} \label{eq:deriv_volume_density}
    \frac{\partial \bar{v}}{\partial D}\bigg|_{D=\bar{D}} = \frac{\partial \bar{v}}{\partial \bar{z}} \bigg(\frac{d\bar{D}}{d\bar{z}}\bigg)^{-1} = \frac{\bar{v}(\bar{D})\gamma}{\bar{D}} \bigg[-4 + \frac{2}{\bchi{} \H{}} + \frac{\H{}'}{\H{}^2} - \frac{\gamma}{a^2 \bar{D}} \frac{d^2\bar{D}}{d\bar{z}^2}\bigg]\,,
\end{equation}
and, finally, plugging Eqs.~\eqref{eq:perturbed_volume_density_final} and~\eqref{eq:deriv_volume_density} into Eq.~\eqref{eq:volume_perturbations_def} we obtain
\begin{equation}\begin{split}
    \frac{V(D, \hat{n}) - \bar{V}(D)}{\bar{V}(D)} = &2\frac{\delta\chi}{\bchi{}} + \bigg(\cot\theta + \frac{\partial}{\partial \theta}\bigg)\delta\theta + \frac{\partial \delta\phi}{\partial \phi} + \bm{v}\cdot\hat{n} - 3\Phi + \frac{d\delta\chi}{d\bchi} - \frac{\gamma}{\bar{D}\H{}}\frac{d\Delta D}{d\bchi} + \\
    &-\gamma \bigg(-4 + \frac{2}{\bchi{} \H{}} + \frac{\H{}'}{\H{}^2} - \frac{\gamma}{a^2 \bar{D}} \frac{d^2\bar{D}}{d\bar{z}^2}\bigg) \frac{\Delta D}{\bar{D}}\,.
\end{split}
\end{equation}

We can further explicitate this expression, by using that the perturbation on the comoving distance is given by (see~\cite{Bonvin2011,Bertacca:2017vod})
\begin{equation} \label{eq:perturbations_chi}
    \delta \chi = \int_0^{\bchi{}} d\chi' (\Phi+\Psi)\,,
\end{equation}
which leads to
\begin{equation}
    \frac{d\delta\chi}{d\bchi{}} = \Phi + \Psi\,,
\end{equation}
while we recognise in the angular term the lensing contribution it its usual form~\cite{Bonvin2011}
\begin{equation}
    \bigg(\cot\theta + \frac{\partial}{\partial \theta}\bigg)\delta\theta + \frac{\partial \delta\phi}{\partial \phi} = - \int_0^{\bchi{}} d\chi' \frac{\bchi{}-\chi'}{\bchi{}\chi'} \nabla_{\Omega}^2 (\Phi + \Psi)\,.
\end{equation}

Lastly, we follow~\cite{Fonseca:2023uay} and expand
\begin{equation} \label{eq:deriv_deltaD}
    \frac{d\Delta D}{d\bchi} = \frac{d}{d\bchi} \bigg(\frac{\Delta D}{\bar{D}} \bar{D}\bigg) =  \bar{D}\frac{d}{d\bchi} \bigg(\frac{\Delta D}{\bar{D}}\bigg) + \frac{\Delta D}{\bar{D}}\frac{d\bar{D}}{d\bchi{}} = \bar{D}\frac{d}{d\bchi} \bigg(\frac{\Delta D}{\bar{D}}\bigg) + \frac{\H{}\bar{D}}{\gamma} \frac{\Delta D}{\bar{D}}\,,
\end{equation}
and finally, combining Eqs.~\eqref{eq:perturbations_chi}-\eqref{eq:deriv_deltaD} we end up with
\begin{equation}\begin{split}
    \frac{V(D, \hat{n}) - \bar{V}(D)}{\bar{V}(D)} = &\frac{2}{\bchi{}} \int_0^{\bchi{}} d\chi' (\Phi+\Psi) - \int_0^{\bchi{}} d\chi' \frac{\bchi{}-\chi'}{\bchi{}\chi'} \nabla_{\Omega}^2 (\Phi + \Psi)  + \bm{v}\cdot\hat{n} - 2\Phi + \Psi \\
    &- \frac{\gamma}{\H{}}\frac{d}{d\bchi} \bigg(\frac{\Delta D}{\bar{D}}\bigg) - \bigg[1 + \gamma \bigg(-4 + \frac{2}{\bchi{} \H{}} + \frac{\H{}'}{\H{}^2} - \frac{\gamma}{a^2 \bar{D}} \frac{d^2\bar{D}}{d\bar{z}^2}\bigg)\bigg] \frac{\Delta D}{\bar{D}}\,.
\end{split}
\end{equation}

Combining the results for both the density and volume perturbations, we ultimately obtain the general expression~\eqref{eq:counts_general} for the number count in luminosity distance space.

\section{Numerical implementation}
\label{app:source_terms}

For the numerical implementation in \texttt{EFTCAMB}, we want to express the 2-point correlation function of Eq.~\eqref{eq:cross_correlations} in a convenient form. We do so following closely the original \texttt{CAMB} Source paper \cite{Challinor2011}. We define a source object $S_{\Delta}(\bm{k})$ as 
\begin{equation}
    S^i_{\Delta}(\bm{k}) \equiv \int_0^{\eta_0} d\eta \, W^i(\eta) \, \hat{\Delta}(\eta, {\bm k}) \, j_{\ell}(k\chi)\,,
\end{equation}
so that Eq.~\eqref{eq:spherical_coefficients_def} becomes
\begin{equation}
    \Delta^{\rm obs} (D^i,\hat{n}) = \sum_{\ell=0} \sum_{m=-\ell}^{\ell} \bigg[ 4\pi i^{\ell} \int \frac{d^3 k} {(2\pi)^{3/2}} Y^*_{\ell m}(\hat{k}) S^i_{\Delta}(\bm{k}) \bigg] Y_{\ell m}(\hat{n})\,,
\end{equation}
and the angular power spectra of Eq.~\eqref{eq:cross_correlations} simply read
\begin{equation}
    C^{ij}_\ell = 4\pi \int d\log k \, \langle S^i_{\Delta}(\bm{k}) S^{j *}_{\Delta}(\bm{k})\rangle\,.
\end{equation}

Recalling Eq.~\eqref{eq:number_count_final}, we re-write $S^i_{\Delta}$, splitting it into a sum of source terms, each corresponding to a specific relativistic correction
\begin{equation}
    S^i_{\Delta}(\bm{k}) = S^i_{\delta}(\bm{k}) + S^i_{\rm D}(\bm{k}) + S^i_{\rm DSD}(\bm{k}) + S^i_{\rm L}(\bm{k}) + S^i_{\rm TD}(\bm{k}) + S^i_{\rm ISW}(\bm{k}) + S^i_{\rm Pot}(\bm{k}) + S^i_{\nabla\Phi}(\bm{k}) + S^i_{\rm SF}(\bm{k})\,.
\end{equation}

Each source term can be derived by performing the expansion~\eqref{eq:spherical_coefficients_def} on each term at the RHS of Eq.~\eqref{eq:number_count_final} and factoring out the integration over time.

\subsection{Derivation of the source terms}

\begin{itemize}
    \item We begin by deriving the source terms corresponding to the perturbations at the source position and that do not contain spatial derivatives. These are $S^i_{\delta}$, $S^i_{\Phi}$, $S^i_{\dot{\Phi}}$, $S^i_{\Psi}$ and $S^i_{\rm SF}$. For a generic term of the form
\begin{equation}
    \Delta_{\hat{s}} (D^i, \hat{n}) = \int_0^{\eta_0} d\eta \, W^i(\eta) \, \mathcal{A}_{\hat{s}}(\eta) \,  \hat{s}(\eta, \hat{n})\,,
\end{equation}
the expansion in spherical harmonics is the same as Eq.~\eqref{eq:spherical_coefficients_def}, so the source term simply reads
\begin{equation}
    S^i_{\hat{s}}(\bm{k}) = \int_0^{\eta_0} d\eta \, \big[ W^i(\eta) \, \mathcal{A}_{\hat{s}}(\eta) \, \hat{s}( \bm{k}, \eta) \big] j_\ell(k\chi)\,,
\end{equation}
giving Eqs.~\eqref{eq:source_density},~\eqref{eq:source_pot} and~\eqref{eq:source_scalarfield}.

\item We then have perturbation terms that are integrated over the GW path, such as the ISW and the Shapiro time-delay terms. These only introduce a small extra complication to the calculation. The terms are in the generic form
\begin{equation}
    \Delta_{\hat{s}}(D^i, \hat{n}) = \int_0^{\infty} d\chi' \, W^i(\chi') \mathcal{A}_{\hat{s}}(\chi') \int_0^{\chi'} d\chi \hat{s}(\chi,\hat{n})\,,
\end{equation}
where $W^i(\chi) \equiv dD/d\chi \, W^i(D)$. Before expanding in Fourier-modes, we want to manipulate these terms changing the order of integration. Doing so, transforms $\Delta_{\hat{s}}$ as
\begin{equation}
    \Delta_{\hat{s}}(D^i, \hat{n}) = \int_0^{\infty} d\chi \int_{\chi}^{\infty} d\chi' W^i(\chi') \mathcal{A}_{\hat{s}}(\chi') \hat{s}(\chi,\hat{n})\,,
\end{equation}
amd switching to integrations over the conformal time, we obtain
\begin{equation}
    \Delta_{\hat{s}}(D^i, \hat{n}) = \int_0^{\eta_0} d\eta \bigg[\int_0^\eta d\eta' W^i(\eta') \mathcal{A}_{\hat{s}}(\eta')\bigg] \hat{s}(\eta,\hat{n})\,,
\end{equation}
where we also have renamed the silent integration variables, switching primed and un-primed $\eta$. Now, $\hat{s}$ is in the proper form to perform the usual expansion in spherical harmonics, which then allows us to read off
\begin{equation}
    S^i_{\hat{s}} (\bm{k}) = \int_0^{\eta_0} d\eta \bigg[\int_0^\eta d\eta' W^i(\eta') \mathcal{A}_{\hat{s}}(\eta')\bigg] \hat{s}(\bm{k}, \eta) j_\ell(k\chi)\,,
\end{equation}
giving Eqs.~\eqref{eq:source_td} and~\eqref{eq:source_isw}. We have momentarily set aside the lensing term, that requires a slightly separate treatment.

\item For the Doppler term instead, we have
\begin{equation}\label{eq:peculiar_velocity}
        \Delta_{D} ({\bm \chi}, \eta) = \int_0^{\eta_0} d\eta \, W^i(\eta) \mathcal{A}_D(\eta) \int \frac{d^3 k}{(2\pi)^{3/2}} e^{i \sprod{k}{\chi}} v_{\parallel}({\bm k}, \eta)\,,
\end{equation}
where we have already performed the expansion in Fourier modes. We now use the definition $v_{\parallel}(\bm{k}, \eta) \equiv i v(\bm{k}, \eta) \sprod{k}{\hat{n}}$, i.e. the velocity field is irrotational. Then, Eq.~\eqref{eq:peculiar_velocity} becomes
\begin{equation}
    \Delta_{D} ({\bm \chi}, \eta) = \int_0^{\eta_0} d\eta \, W^i(\eta) \mathcal{A}_D(\eta) \int \frac{d^3 k}{(2\pi)^{3/2}} \, v(\bm{k}, \eta) \, d_{k\chi} e^{i \sprod{k}{\chi}}\,,
\end{equation}
and further expanding the exponential, we obtain
\begin{equation}
    \Delta^D_{\ell m} ({\bm \chi}, \eta) = 4\pi i^{\ell} \int \frac{d^3 k} {(2\pi)^{3/2}} Y^*_{\ell m}(\hat{k}) \int_0^{\eta_0} d\eta \, W^i(\eta)\, \mathcal{A}_D(\eta) \, v(\bm{k}, \eta) \, d_{k\chi} j_{\ell}(k\chi)\,.
\end{equation}

Lastly, we integrate by parts to remove the derivative on the Bessel function. To do so, we use that $d_{k\chi} = (1/k) d/d{\chi} = - (1/k) d/d{\eta}$ and work under the assumption that the surface terms can be put to 0, as the window functions goes to 0 at both ends. We obtain
\begin{equation}
    \Delta^D_{\ell m} ({\bm \chi}, \eta) = 4\pi i^{\ell} \int \frac{d^3 k} {(2\pi)^{3/2}} Y^*_{\ell m}(\hat{k}) \int_0^{\eta_0} d\eta \frac{d}{d\eta} \bigg[\frac{W^i(\eta)}{k}\, \mathcal{A}_D(\eta) \, v(\bm{k}, \eta) \bigg] \, j_{\ell}(k\chi)\,,
\end{equation}
from which we read Eq.~\eqref{eq:source_doppler} for $S^i_D(\bm{k})$.

\item We take a similar approach for the DSD term, which contains the derivative of the peculiar velocity along the null geodesics. For this term we have
\begin{equation}\begin{split}
    \Delta_{\rm DSD} &=\int_0^{\eta_0} d\eta \, W^i(\eta) \mathcal{A}_{\rm DSD}(\eta) \, \sprod{\hat{n}}{\nabla} \big(v_{\parallel}(\bm{\chi}, \eta)\big) =\\ &=\int_0^{\eta_0} d\eta \, W^i(\eta) \mathcal{A}_{\rm DSD}(\eta) \int \frac{d^3 k}{(2\pi)^{3/2}} \sprod{\hat{n}}{\nabla} \bigg( e^{i \sprod{k}{\chi}} v_{\parallel}({\bm k}, \eta) \bigg) =\\
    &= \int_0^{\eta_0} d\eta \, W^i(\eta) \mathcal{A}_{\rm DSD}(\eta) \int \frac{d^3 k}{(2\pi)^{3/2}} \, v({\bm k}, \eta) \, k \, d^2_{k\chi} e^{i \sprod{k}{\chi}} \,,
\end{split}
\end{equation}
and for the spherical harmonics coefficients
\begin{equation}
    \Delta^{\rm DSD}_{\ell m} = 4\pi i^{\ell} \int \frac{d^3 k} {(2\pi)^{3/2}} Y^*_{\ell m}(\hat{k}) \int_0^{\eta_0} d\eta W^i(\eta)\, \mathcal{A}_{\rm DSD}(\eta) \, v(\bm{k}, \eta) \, k \, d^2_{k\chi} j_{\ell}(k\chi)\,.
\end{equation}

Again, we integrate by parts twice to remove the double derivative on the Bessel function and finally get
\begin{equation}
    \Delta^{\rm DSD}_{\ell m} = 4\pi i^{\ell} \int \frac{d^3 k} {(2\pi)^{3/2}} Y^*_{\ell m}(\hat{k}) \int_0^{\eta_0} d\eta \frac{d^2}{d\eta^2}\bigg[ \frac{W^i(\eta)}{k}\, \mathcal{A}_{\rm DSD}(\eta) \, v(\bm{k}, \eta) \bigg] \, j_{\ell}(k\chi)\,,
\end{equation}
from which we read off Eq.~\eqref{eq:source_lsd} for $S^i_{\rm DSD}(\bm{k})$.

\item The last term containing a derivative along the wave path is the potential term proportional to $\sprod{\hat{n}}{\nabla} \Phi$. For this term, we have
\begin{equation}\begin{split}
    \Delta_{\nabla\Phi}(\bm{\chi}, \eta) &= \int_0^{\eta_0} d\eta \, W^i(\eta) \mathcal{A}_{\rm DSD}(\eta) \, \int \frac{d^3 k}{(2\pi)^{3/2}} \sprod{\hat{n}}{\nabla} \bigg( e^{i \sprod{k}{\chi}} \Phi({\bm k}, \eta) \bigg) =\\
    &= \int_0^{\eta_0} d\eta \, W^i(\eta) \mathcal{A}_{\rm DSD}(\eta) \, \int \frac{d^3 k}{(2\pi)^{3/2}} \, \Phi({\bm k}, \eta) \, k \, d_{k\chi} e^{i \sprod{k}{\chi}}\,,
\end{split}
\end{equation}
yielding
\begin{equation}
    \Delta^{\nabla\Phi}_{\ell m} = 4\pi i^{\ell} \int \frac{d^3 k} {(2\pi)^{3/2}} Y^*_{\ell m}(\hat{k}) \int_0^{\eta_0} d\eta \frac{d}{d\eta}\bigg[ W^i(\eta)\, \mathcal{A}_{\nabla\Phi}(\eta) \, \Phi(\bm{k}, \eta) \bigg] \, j_{\ell}(k\chi)\,.
\end{equation}

Once integrated by parts, that gives $S^i_{\nabla\Phi}(\bm{k})$ (Eq.~\eqref{eq:source_gradphi}).

\item Finally, we look at the lensing therm. First, we perform the same trick implemented above for the ISW and time delay terms, and exchange the order of integration, obtaining
\begin{equation}\label{eq:lensing}
    \Delta_{L}(D^i, \hat{n}) = \int_0^{\infty} d\chi \int_{\chi}^{\infty} d\chi' W^i(\chi') \mathcal{A}_{L}(\chi') \nabla^2_{\Omega} \big[\Phi(\chi,\hat{n})+\Psi(\chi,\hat{n})\big]\,.
\end{equation}

For this term we will work, as it is customary, in the flat sky approximation. Then, instead of the expansion of Eq.~\eqref{eq:spherical_coefficients_def}, we take directly the 2D Fourier transform of the 2D laplacian, that simply brings down a factor $- \ell(\ell+1)$, giving
\begin{equation}
    \Delta^{L}_{\rm \ell m}(D^i, \hat{n}) = - 4\pi i^{\ell} \, \ell(\ell+1) \int \frac{d^3 k} {(2\pi)^{3/2}} Y^*_{\ell m}(\hat{k}) \int_0^{\eta_0} d\eta \int_{0}^{\eta} d\eta' W^i(\eta') \mathcal{A}_{L}(\eta') \big[\Phi(\bm{k}, \eta)+\Psi(\bm{k}, \eta)\big]\,,
\end{equation}
from which follows Eq.~\eqref{eq:source_lensing}.
\end{itemize}

In summary, the final expression for the source terms implemented in \texttt{EFTCAMB} reads
\begin{subequations}
\begin{align}
    S^i_{\delta}(\bm{k}) &= \int_0^{\eta_0} d\eta \bigg[ W^i (\eta) \delta_{\rm gw} (\bm{k}, \eta)\bigg] j_{\ell}(k\chi) \,,\label{eq:source_density} \\[10pt]
    S^i_{D}(\bm{k}) &= \int_0^{\eta_0} d\eta \frac{d}{d\eta} \bigg[\frac{W^i(\eta)}{k}\, \bigg(1 -2\gamma - 2 \big(\beta+1\big)\bigg) v(\bm{k}, \eta) \bigg] \, j_{\ell}(k\chi) \,,\label{eq:source_doppler} \\[10pt]
    S^i_{\rm DSD}(\bm{k}) &= \int_0^{\eta_0} d\eta \frac{d^2}{d\eta^2}\bigg[ \frac{W^i(\eta)}{k}\, 2 \frac{\gamma}{\H} \, v(\bm{k}, \eta) \bigg] \, j_{\ell}(k\chi) \,,\label{eq:source_lsd} \\[10pt]
    \begin{split}
    S^i_{L}(\bm{k}) &= - \ell(\ell+1) \int_0^{\eta_0} d\eta \int_0^\eta d\eta' W(\eta') \times \\
    &\times \bigg[\bigg(\frac{\beta-1}{2}\bigg) \frac{\chi' - \chi}{\chi\chi'} + \frac{\gamma}{2\H\chi'^2}\bigg] \big[\Phi(\bm{k}, \eta) + \Psi(\bm{k}, \eta)\big] j_{\ell}(k\chi)\,,\label{eq:source_lensing} \end{split}\\[10pt]
    S^i_{\rm TD}(\bm{k}) &= \int_0^{\eta_0} d\eta \bigg[\int_0^\eta d\eta' W(\eta') \bigg(\frac{1-\beta}{\chi'} + \frac{\gamma}{\H\chi'^2}\bigg) \bigg] \big[\Phi(\bm{k}, \eta) + \Psi(\bm{k}, \eta)\big] j_{\ell}(k\chi)\,,\label{eq:source_td} \\[10pt]
    S^i_{\rm ISW}(\bm{k}) &= \int_0^{\eta_0} d\eta \bigg[\int_0^\eta d\eta' W(\eta') \, 2 \big(\beta + 1\big)\bigg] \big[\dot{\Phi}(\bm{k}, \eta) + \dot{\Psi}(\bm{k}, \eta)\big] j_{\ell}(k\chi) \,,\label{eq:source_isw} \\[10pt]
    S^i_{\rm Pot}(\bm{k}) &= \int_0^{\eta_0} d\eta \, W^i(\eta) \bigg[ \bigg(\beta - 1 - \frac{\gamma}{\bchi\H}\bigg) \Phi + \frac{\gamma}{\H} \dot{\Phi} + \bigg(1-\frac{\gamma}{\chi\H} + 2 \big(\beta + 1\big)\bigg) \Psi \bigg] j_{\ell}(k\chi)\,, \label{eq:source_pot} \\[10pt]
    S^i_{\nabla \phi}(\bm{k}) &= \int_0^{\eta_0} d\eta \frac{d}{d\eta}\bigg[ W^i(\eta)\, \frac{\gamma}{\H} \, \Phi(\bm{k}, \eta) \bigg] \, j_{\ell}(k\chi)\,, \label{eq:source_gradphi} \\[10pt]
    S^i_{\rm SF}(\bm{k}) &= \int_0^{\eta_0} d\eta \, W^i(\eta) \bigg\{\gamma \frac{\alpha_M}{2} \dot{\bigg(\frac{\delta \varphi}{\varphi}\bigg)} + \bigg[\gamma \frac{\dot{\alpha_M}}{2} + \frac{\alpha_M}{2} \bigg(\gamma \frac{\dot{\H}}{\H}-\beta - 1\bigg) \bigg] \frac{\delta \varphi}{\varphi} \bigg\} j_{\ell}(k\chi)\,. \label{eq:source_scalarfield}
\end{align}
\end{subequations}
with
\begin{equation}
   \chi = (\eta_0 - \eta) \quad \text{and} \quad W^i(\eta) = \frac{\sqrt{F} (\eta_0 - \eta) \H}{a} \bigg[1+ \frac{1}{(\eta_0-\eta) \H} - \frac{\alpha_M}{2}\bigg] \, W^i\big(D(\eta)\big)\,.
\end{equation}

We note that our Eq.~\eqref{eq:source_lensing} differs from that of~\cite{Fonseca:2023uay} not only because of the amplitude entering the square bracket - as discussed in the text - but also by the overall minus sign, which is instead consistent with~\cite{Bonvin2011, Challinor2011}. This does not affect the lensing auto-correlation, but it impacts all cross-correlations of lensing with the other relativistic effects.

\bibliographystyle{JHEP}
\bibliography{references.bib}

\end{document}